%% file: main.tex
\documentclass[sigconf]{acmart}

\copyrightyear{2022} 
\acmYear{2022} 
\setcopyright{acmlicensed}
\acmConference[ICSE '22]{44th International Conference on Software Engineering}{May 21--29, 2022}{Pittsburgh, PA, USA}
\acmBooktitle{44th International Conference on Software Engineering (ICSE '22), May 21--29, 2022, Pittsburgh, PA, USA}
\acmPrice{15.00}
\acmDOI{10.1145/3510003.3510202}
\acmISBN{978-1-4503-9221-1/22/05}

\usepackage{amsmath,amsfonts}

\usepackage{hyperref}
\usepackage{subcaption}
\newsavebox{\mybox}

\input{header}

\begin{document}

\title{Fairness-aware Configuration of Machine Learning Libraries}

\author{Saeid Tizpaz-Niari}
\email{saeid@utep.edu}
\affiliation{%
  \country{University of Texas at El Paso}
}

\author{Ashish Kumar}
\email{azk640@psu.edu}
\affiliation{%
  \country{Pennsylvania State University}
}

\author{Gang Tan}
\email{gtan@psu.edu}
\affiliation{%
  \country{Pennsylvania State University}
}

\author{Ashutosh Trivedi}
\email{ashutosh.trivedi@colorado.edu}
\affiliation{%
  \country{University of Colorado Boulder}
}

\begin{abstract}
  This paper investigates the parameter space of machine learning (ML)
  algorithms in aggravating or mitigating fairness bugs. Data-driven
  software is increasingly applied in social-critical applications
  where ensuring fairness is of
  paramount importance.
  The existing approaches focus on addressing fairness bugs
  by either modifying the input dataset
  or modifying the learning algorithms.
  On the other hand, the selection of hyperparameters, which provide finer controls of ML algorithms,
  may enable a less intrusive approach
  to influence the fairness.
  \emph{Can hyperparameters amplify or suppress discrimination present
  in the input dataset?}
  \emph{How can we help programmers in detecting, understanding, and exploiting
  the role of hyperparameters to improve the fairness?}

  We design three search-based software testing algorithms to
  uncover the precision-fairness frontier of the hyperparameter space.
  We complement these algorithms with statistical
  debugging to explain the role of these parameters in improving fairness.
  We implement the proposed approaches in the tool \toolname{}
  (PARameter FAIrness Testing for ML Libraries) and show its effectiveness and
  utility over five mature ML algorithms as used in six social-critical
  applications. In these applications, our approach successfully identified
  hyperparameters that significantly improve (vis-a-vis the state-of-the-art
  techniques) the fairness without sacrificing precision. Surprisingly,
  for some algorithms (e.g., random forest), our approach showed that certain
  configuration of hyperparameters (e.g., restricting the search space of attributes)
  can amplify biases across applications. Upon further investigation,
  we found intuitive explanations of these phenomena, and the results corroborate
  similar observations from the literature.
\end{abstract}

\maketitle

\section{Introduction}
\label{sec:introduction}
\input{introduction}

\section{Background}
\label{sec:background}
\input{background}

\section{Overview}
\label{sec:overview}
\input{overview}

\section{Problem Definition}
\label{sec:problem}
\input{definition}

\section{Approach}
\label{sec:approach}
\input{approach}

\section{Experiment}
\label{sec:expr}
\input{experiments}

\section{Discussion}
\label{sec:discussion}
\input{discussion}

\section{Conclusion}
\label{sec:conclusion}
\input{conclusion}

\input{ack}

\bibliographystyle{ACM-Reference-Format}
\bibliography{papers}

\end{document}

%% file: header.tex
\usepackage{amsthm}
\usepackage{mdframed}
\usepackage{centernot}
\usepackage{comment}

\newcommand{\Aa}{\mathcal{A}}

\newcommand{\Ff}{\mathcal{F}}

\newcommand{\Hh}{\mathcal{H}}

\newcommand{\Xx}{\mathcal{X}}
\newcommand{\Yy}{\mathcal{Y}}

\newcommand{\Dd}{\mathcal{D}}

\newcommand\vd[2]{d_{i, p}}
\newcommand{\set}[1]{\left\{ #1 \right\}}

\newcommand{\toolname}{\textsc{Parfait-ML}\xspace}
\usepackage{tikz}

\newtheorem{definition}{Definition}[section]

\definecolor{gold}{rgb}{0.99,0.78,0.07}
\usetikzlibrary{arrows,shapes,snakes,automata,backgrounds,positioning,decorations.pathmorphing,
	decorations.markings,calc}

\tikzstyle{dtreenode}=[draw=blue!10!gray,rounded rectangle, minimum size=5mm,fill=blue!10!white]
\tikzstyle{dtreeleaf}=[draw=black!60,minimum width=1cm,minimum height=0.4cm,rectangle,fill=blue!50!white]
\tikzset{every loop/.style={looseness=7}}
\tikzset{
	gluon/.style={decorate,draw=black,
		decoration={coil,amplitude=1pt, segment length=5pt}}
}
\tikzset{
	gluon1/.style={decorate,draw=black,
		decoration={coil,amplitude=3pt, segment length=3pt}}
}
\tikzset{
	gluonew/.style={decorate,draw=black,
		decoration={coil,amplitude=1pt, segment length=2pt}}
}

\tikzset{bicolor/.style args={#1 and #2 and #3}{
		path picture={
			\tikzset{rounded corners=0}
			\fill [#1] (path picture bounding box.south west)
			rectangle
			($(path picture  bounding box.north west)!#3!(path picture bounding
			box.north east)$);
			\fill [#2]
			($(path picture bounding box.south west)!#3!(path picture bounding
			box.south east)$)
			rectangle (path picture bounding box.north east);
}}}

\tikzset{tricolor/.style args={#1 and #2 and #3 and #4 and #5}{
		path picture={
			\tikzset{rounded corners=0}
			\fill [#1] (path picture bounding box.south west)
			rectangle
			($(path picture  bounding box.north west)!#4!(path picture bounding
			box.north east)$);
			\fill [#2]
			($(path picture bounding box.south west)!#4!(path picture bounding
			box.south east)$)
			rectangle
			($(path picture  bounding box.north west)!#5!(path picture bounding
			box.north east)$);
			\fill [#3]
			($(path picture bounding box.south west)!#5!(path picture bounding
			box.south east)$)
			rectangle (path picture bounding box.north east);
}}}

\usepackage[many]{tcolorbox}
\tcbuselibrary{listings,skins}

\lstdefinestyle{mystyle}{
  xleftmargin=0pt,
   basicstyle={\footnotesize\ttfamily},
   aboveskip=3mm,
   belowskip=3mm,
   keywordstyle=\bfseries,
   showstringspaces=false,
  escapechar=?,
  language=Java
}
\definecolor{code_indent}{HTML}{CCCCCC}

\newtcblisting{mylisting}[2][]{
  arc=0pt, outer arc=0pt,
  listing only,
  listing style=mystyle,
  title={\large #2},
  #1
}

 \definecolor{dkgreen}{rgb}{0,0.6,0}
 \definecolor{gray}{rgb}{0.5,0.5,0.5}
 \definecolor{mauve}{rgb}{0.58,0,0.82}

\definecolor{cadmiumgreen}{rgb}{0.0, 0.42, 0.24}
\definecolor{verde}{rgb}{0.25,0.5,0.35}
\definecolor{jpurple}{rgb}{0.5,0,0.35}
\definecolor{darkgreen}{rgb}{0.0, 0.2, 0.13}

\usepackage[ruled,vlined,linesnumbered,lined]{algorithm2e}
\usepackage{algpseudocode}

\usepackage{mathtools,xparse}

\usepackage{tabu}
\usepackage{multirow}

\usepackage{pifont}
\usepackage{enumerate}
\usepackage[shortlabels]{enumitem}

\usepackage{pgfplotstable}
\usepackage{pgfplots}

\usepackage[noframe]{showframe}
\usepackage{framed}

 {\endMakeFramed}
 \definecolor{shadecolor}{gray}{0.85}

\definecolor{bgblue}{RGB}{245,243,253}
\definecolor{ttblue}{RGB}{91,194,224}

\mdfdefinestyle{mystyle}{%
  rightline=true,
  innerleftmargin=10,
  innerrightmargin=10,
  outerlinewidth=3pt,
  topline=false,
  rightline=false,
  bottomline=false,
  skipabove=\topsep,
  skipbelow=false
}

\newtcolorbox{myboxi}[1][]{
  breakable,
  title=#1,
  colback=white,
  colbacktitle=white,
  coltitle=black,
  fonttitle=\bfseries,
  bottomrule=0pt,
  toprule=0pt,
  leftrule=3pt,
  rightrule=3pt,
  titlerule=0pt,
  arc=0pt,
  outer arc=0pt,
  colframe=black!50,
}

\newtcolorbox{myboxii}[1][style=mystyle]{
  breakable,
  freelance,
  colback=white,
  colbacktitle=white,
  coltitle=black,
  fonttitle=\bfseries,
  bottomrule=0pt,
  boxrule=0pt,
  colframe=white,
  after skip=0pt,
  overlay unbroken and first={
    \draw[white!75!black,line width=3pt]
    ([yshift=-9pt]frame.north west) --
    ([yshift=9pt]frame.south west);
  },
  }

%% file: introduction.tex
Data-driven software applications are an integral part of modern life
impacting every aspect of societal structure, ranging from education and health
care to criminal justice and finance~\cite{compas-software,deloitte-software}.
Since these algorithms learn from prior experiences, it is not
surprising that they encode historical and present biases due to displacement,
exclusion, segregation, and injustice.
The resulting software may particularly disadvantage minorities
and protected groups
\footnote{
A wall street journal article showed Deloitte,
a life-insurance risk assessment
software can discriminate based on the protected health status of
applicants~\cite{WSJ-insurace,dwork2012fairness}.
FICO, a credit risk
assessment software, is found to predict higher risks for black
non-defaulters~\cite{hardt2016equality} than white/Asian ones.
COMPAS risk assessment software in criminal justice is shown to predict higher
risks for black defendants~\cite{compas-article}.}
and be found non-compliant with law such as the US Civil
Rights Act~\cite{blumrosen1978wage}. Therefore,
helping programmers detect and mitigate fairness bugs
in social-critical data-driven software systems is crucial to
ensure inclusion in our modern, increasingly digital society.

The software engineering (SE) community has invested substantial efforts
to improve the fairness of ML software~\cite{FairnessTesting,
10.1145/3338906.3338937,udeshi2018automated,ADF,chakraborty2020fairway}.
Fairness has been treated as a critical meta-properties
that requires an analysis beyond functional correctness
and measurements such as prediction accuracy~\cite{10.1145/3236024.3264838}.
However, the majority of previous work within the SE community evaluates fairness on the
\textit{ML models}
after training~\cite{FairnessTesting,udeshi2018automated,10.1145/3338906.3338937,ADF},
while the programmer supports to improve fairness during the inference of
models (i.e., training process) is largely lacking.
%
%
%

The role of training processes in amplifying or suppressing vulnerabilities and bugs
in the input dataset is well-documented~\cite{zhang2020machine}. The training
process typically involves tuning of \textit{hyperparameters}: variables
that characterize the hypothesis space of ML models and define a trade-off
between complexity and performance.
Some prominent examples of hyperparameters include {\tt l1} vs.\ {\tt l2} loss
function in support vector machines, the maximum depth of a decision tree,
and the number of layers/neurons in deep neural networks.
Hyperparameters are crucially different
from the ML model parameters in that they cannot be learned from the input dataset alone.
In this paper, we investigate the impact of hyperparameters on ML fairness
and propose a programmer support system to develop fair data-driven software.

We pose the following research questions:
\emph{To what extent can hyperparameters influence the biases present
in the input dataset?}
\emph{Can we assist ML library developers in identifying and explaining
fairness bugs in the hyperparameter space?}
\emph{Can we help ML users to exploit the hyperparamters
to imporive fairness?}

We present \toolname (PARameter FAIrness Testing for ML Libraries): a
search-based testing and statistical debugging framework that supports
ML library developers and users to detect, understand, and exploit configurations
of hyperparameters to improve ML fairness without impacting functionality.
%
%
We design and implement three dynamic search algorithms
(\textit{independently random}, \textit{black-box evolutionary}, and \textit{gray-box
evolutionary}) to find configurations that simultaneously maximize and minimize group-based
fairness with a constraint on the prediction accuracy. Then, we leverage
statistical learning methods~\cite{alhazen,DBLP:conf/aaai/Tizpaz-NiariCCT18} to explain what hyperparameters distinguish
low-bias models from high-bias ones. Such explanatory models specifically aid
ML library maintainers to localize a fairness bug. Finally, we show that \toolname can
effectively (vis-a-vis the state-of-the-art
techniques) aid ML users to find a configuration that mitigates bias
without degrading the prediction accuracy.

We evaluate our approach on five well-established machine learning
algorithms over six fairness-sensitive training tasks. Our results show
that for some algorithms, there are hyperparameters that consistently impact
fairness across different training tasks. For example, \texttt{max\_feature} parameter
in random forest can aggravate the biases for some of its values such as $\log_2$(\#num. features)
beyond a specific training task. These observations corroborate similar empirical observations
made in the literature~\cite{DBLP:conf/icse/ZhangH21}.

In summary, the key contributions of this paper are:
\begin{enumerate}
  \item the first approach to support \textit{ML library
  maintainers} to understand the fairness implications of algorithmic configurations;
  \item three search-based algorithms to approximate
  the Pareto curve of hyperparameters against the fairness and accuracy;
  \item a \textit{statistical debugging} approach to localize
  parameters that systematically influence fairness in five popular and
  well-establish ML algorithms over six fairness-critical datasets;
  \item a \textit{mitigation} approach to effectively find configurations that
  reduce the biases (vis-a-vis the state-of-the-art); and
  \item an implementation of \toolname (PARameter FAIrness
  Testing for ML Libraries) and its experimental evaluation on multiple
  applications, available at: \url{https://github.com/Tizpaz/Parfait-ML}.
\end{enumerate}

%% file: background.tex
\noindent \textbf{Fairness Terminology and Measures.}
\label{subsec:Fairness-Terminology}
Let us first recall some fairness vocabulary.
We consider \textit{binary classification} tasks where a class label
is \textit{favorable} if it gives a benefit to an individual such as low credit
risk for loan applications (default), low reoffend risk for parole assessments
(recidivism), and high qualification score for job hiring.
Each dataset consists of a number of \textit{attributes} (such as income,
employment status, previous arrests, sex, and race) and a set
of \textit{instances} that describe the value of attributes for each individual.
We assume that each attribute is labeled as \textit{protected} or \textit{non-protected}.
According to ethical and legal requirements, ML software should not \textit{discriminate}
on the basis of an individual's protected attributes such as sex, race, age, disability,
colour, creed, national origin, religion, genetic information, marital status, and sexual orientation.

There are several well-motivated characterizations of fairness.
Fairness through unawareness (FTU)~\cite{dwork2012fairness} requires masking
protected attributes during training. However,
$FTU$ is not effective since protected and non-protected attributes
often correlate (e.g., ZIP code and race), and biases are introduced from non-protected attributes.
Fairness through awareness (FTA)~\cite{dwork2012fairness} is
an \textit{individual fairness} notion that requires that
two \textit{individuals} with similar non-protected attributes
are treated equally.

\textit{Group fairness} requires the statistics of ML outcomes for different
\emph{protected groups} to be similar~\cite{hardt2016equality}.
There are multiple metrics to measure group fairness
in ML software. Among them, \textit{equal opportunity difference} (EOD) measures
the difference between the true positive rates (TPR) of two protected groups.
Similarly, \textit{average odd difference} (AOD) is the average of differences
between the true positive rates (TPR) and the false positive rates (FPR) of
two protected
groups~\cite{bellamy2019ai,chakraborty2020fairway,DBLP:conf/icse/ZhangH21}. These
metrics can naturally be generalized to handle situations where protected
attributes may have more than two values.
For instance, if race is a protected attribute, then the EOD
is the maximum EOD among any two race groups. \textit{This paper focuses
on group fairness}.
\vspace{0.5em}

\label{subsec:ML-system}
\begin{figure}[!t]
    \centering
    \includegraphics[width=0.49\textwidth]{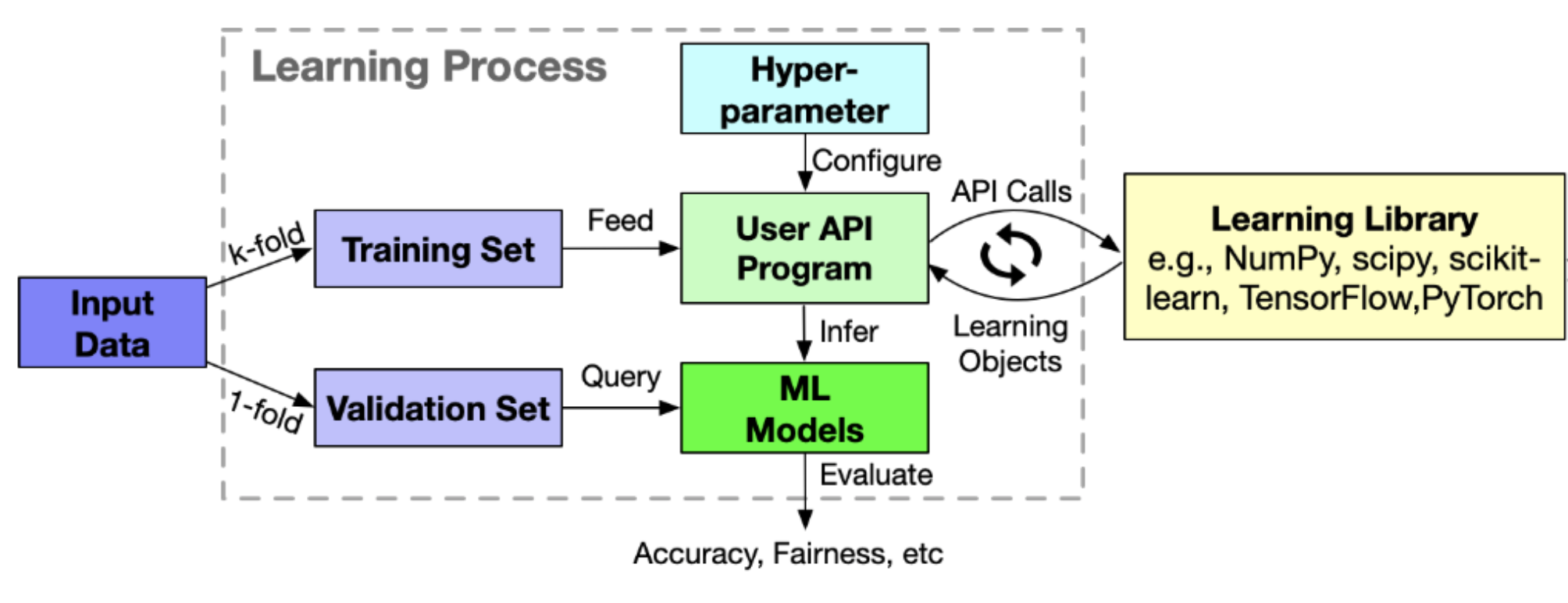}
    \caption{Data-Driven Software System Developments}
    \label{fig:background}
    \vspace{-1.0em}
\end{figure}

\noindent \textbf{Data-Driven Software Systems.}
Data-driven software is distinguished from common software
in that they largely learn their decision logic from datasets.
Consequently, while the traditional software developers explicitly encode decision
logic via control and data structures, the ML programmers and users
provide input data, perform some pre-processing, choose ML algorithms, and tune
hyperparameters to enable data-driven systems to infer a model that encodes
the decision logic.

Figure~\ref{fig:background} shows the key components of a data-driven system.
At a high-level, a data-driven system consists of three major components: input data, a
learning (training) process, and a library framework.
The ML users often provide input data and build an ML model using a programming
interface. The interface interacts with the core ML library (e.g., scikit-learn,
TensorFlow, etc) and constructs different instances of \textit{learning
algorithms} using \textit{hyperparameters}.
Then, they feed the training data into the constructed learning objects to infer the
parameters of an ML model.

As a part of the training process, ML users query the ML model with the \textit{validation set} to
evaluate functional metrics such as prediction accuracy and non-functional
metrics such as EOD for group fairness. At the heart of the learning process,
tuning hyperparameters is particularly challenging since they cannot be estimated
from the input data, and there is no analytical formula to calculate an
appropriate value~\cite{kuhn2013applied}.
We distinguish \textit{algorithm parameters} (i.e., hyperparameters) such as tolerance of
optimization in SVMs, maximum features to search in random forest, and minimum samples in
leaf nodes of decision trees that set before training from \textit{model parameters} that
are inferred automatically after training such as the split feature of
decision tree nodes, the weights of neurons in neural networks, and coefficients
of support vector machines.

\vspace{0.5em}
\noindent \textbf{Related Work.}
\label{subsec:related-work}
\noindent \textit{1) Evaluating fairness of ML models.}
\textsc{Themis}~\cite{FairnessTesting} presents a causal testing approach to
measure group discrimination on the basis of protected attributes.
Particularly, they measure the difference between the fairness metric of two subgroups
by \textit{counterfactual} queries; i.e., they sample inputs where the protected attributes
are A and compare the fairness to a counterfactual scenario where the protected attributes
are set to B.
Agarwal et al.~\cite{10.1145/3338906.3338937} present a black-box testing
technique to detect individual discrimination: two people with similar
features other than protected ones receive different ML predictions. They approximate
ML models with decision trees and use symbolic execution techniques over the tree structure
to find discriminatory instances.
Udeshi et al.~\cite{udeshi2018automated} present a two-step technique
that first uniformly and randomly
search the input data space to find a discriminatory instance and then
locally perturb those instances to further generate biased test cases.
Adversarial discrimination finder~\cite{ADF}
is an adversarial training method to generate individual discrimination
instances in deep neural networks.
These works focus on testing individual ML models and improving their fairness.
We focus on ML libraries and study how algorithm configurations impact fairness.

\noindent \textit{2) Inprocessing methods for bias reduction.}
A body of work considers inprocess algorithms to mitigate biases in ML
predictions. Adversarial debiasing~\cite{zhang2018mitigating} is a technique
based on adversarial learning to infer a classifier that maximizes
the prediction accuracy and simultaneously minimizes adversaries' capabilities to guess the
protected attribute from the ML predictions. Prejudice remover~\cite{6413831} adds
a fairness-aware regularization term to the learning objective and minimizes
the accuracy and fairness loss. This line of work requires the modification
of learning algorithms either in the loss function or the parameter of ML models.
Exponentiated gradient~\cite{agarwal2018reductions} is a
meta-learning algorithm to mitigate biases. The approach infers a family
of classifiers to maximize prediction
accuracy subject to fairness constraints. Since this approach
assumes black-box access to the learning algorithms,
we evaluate the effectiveness of \toolname in \textit{mitigating biases} with this
baseline (see Subsection~\ref{sec:rq4}).

\noindent \textit{3) Combining pre-processing and inprocessing bias reductions.}
\textsc{Fairway}~\cite{chakraborty2020fairway} uses a two step mitigation approach.
While pre-processing, the dataset is divided into privileged and unprivileged
groups where the respective ML models train independently from one another.
Then, they compare the prediction outcomes for the same instance to find and remove discriminatory samples.
Given the pre-processed dataset,
the inprocess step uses a multi-objective optimization (\textsc{FLASH})~\cite{8469102}
to find an algorithm configuration that maximizes both
accuracy and fairness. The work
focuses on using hyperparameters to \textit{mitigate} biases in a subset of
hyperparameters and a limited number of
algorithms. In particular, they require a careful selection of relevant hyperparameters.
Our approach, however, does not require a manual selection of hyperparameters.
Instead, our experiments show that the evolutionary search is effective
in identifying and exploiting fairness-relevant hyperparameters automatically.
In addition, our approach \textit{explains} what hyperparameters influence
fairness.
Such explanatory models can also pinpoint whether some configurations systematically
influence fairness, which can be useful for \textsc{Fairway} to carefully
select a subset of hyperparameters in its search. To show the effectiveness
of \toolname in reducing biases, we compare our approach to \textsc{Fairway}~\cite{chakraborty2020fairway,chakraborty2019software}
(see Subsection~\ref{sec:rq4}).

%% file: overview.tex
We use the example of random forest ensemble~\cite{random-forest}
to overview how \toolname assists ML developers and users
to discover, explain, and mitigate fairness bugs by tuning
the hyperparameters.

\input{overview-Figure}
\vspace{0.5em}
\noindent\textbf{Dataset.}
Adult Census Income~\cite{Dua:2019-census} is a binary classification dataset
that predicts whether an individual has an income over $50K$ a year. The
dataset has $48,842$ instances and $14$ attributes.
For this overview, we start by considering \textit{sex} as the protected attribute.

\vspace{0.5em}
\noindent\textbf{Learning Algorithm.}
Random forest ensemble is a meta estimator that fits a number
of decision trees and uses the averaged outcomes of trees to predict labels.
The ensemble method includes $18$ parameters.
Three parameters are boolean, two are categorical, six are integer,
and seven are real variables. Examples of these parameters are
the maximum depth of the tree, the number of estimator trees, and the minimum
impurity to split a node further.

\vspace{0.5em}
\noindent\textbf{Fairness and Accuracy Criterion.}
We randomly divide the dataset into $4$-folds and use $75\%$ of the dataset
as the training set and $25$\% as the validation set. We measure both accuracy and fairness
metrics after training on ML models using the validation set.
We report the average odd difference ($AOD$) as well as the equal opportunity
difference ($EOD$), which were introduced in Background Section~\ref{sec:background}.
Our accuracy metric is standard: the fraction of correct predictions.

\vspace{0.5em}
\noindent\textbf{Test Cases.} Our approach has three options for generating
test cases: independently random, black-box mutations, and gray-box mutations.
In this section, we use the black-box mutations (see RQ2 in Section~\ref{sec:rq2}).
We run the experiment $10$ times, each for $4$ hours (the default number
of repetition and time-out). We obtain an average of $603$ valid test cases
over $10$ runs.
Since the default parameters of random forests achieve an accuracy of $84\%$,
a valid test case achieves similar or better accuracy. To allow finding a
fair configuration in cases where the default configurations are the most
accurate model, we tolerate $1\%$ accuracy degradations.
The overall accuracy of ML models over the entire corpus of test cases
varies from $83\%$ to $85.7\%$.
Each test case includes the valuation of $18$ algorithm parameters,
accuracy, $AOD$, and $EOD$.

\vspace{0.5em}
\noindent\textbf{Magnitude of Biases.} We report the magnitude
of group biases in the hyperparameter space of random forests.
The minimum and maximum $AOD$ are $5.8\% (+/- 0.6\%)$
and $19.0\% (+/- 0.4\%)$, respectively. The values are the average of $10$ runs,
with $95\%$ confidence interval reported in the
parenthesis, and higher values indicate stronger
biases. Similarly, the minimum and maximum $EOD$ are $4.8\% (+/- 0.4)$
and $32.3\% (+/- 0.6)$. These results show that within $2.7\%$ accuracy margins,
there can be over $13\%$ and $27\%$ difference in $AOD$ and $EOD$, respectively.
These results indicate that different configurations of random forests
can indeed amplify or suppress the biases from the input dataset.
The details of relevant experiments for different datasets and
learning algorithms are reported in RQ1 (Section~\ref{sec:rq1}).

\vspace{0.5em}
\noindent\textbf{Explanation of Biases.}
Our next goal is to explain what configurations of hyperparameters
influence group-based fairness.

\noindent\textit{Clustering.} We first partition generated test cases in
the domain of fairness ($AOD$) versus accuracy. Particularly,
we apply the Spectral clustering algorithm where the number of
partitions are set to three.
Figure~\ref{fig:overview-TreeRegressor} (a) shows the three clusters
identified from the generated test cases. Looking into the figure, we
see that green and orange clusters have similar accuracy; however,
they have significantly different biases ($AOD$). Additionally, the blue cluster
achieves better accuracy with a similar $AOD$ to the green cluster.

\noindent\textit{Tree Classifiers.}
Next, we use CART tree classifiers~\cite{Breiman/1984/CART} to
explain the differences between the clusters in terms of algorithm
parameters as shown in Figure~\ref{fig:overview-TreeRegressor} (b).
Each node in the tree shows a split parameter, the number of samples
reaching the node, and sample distributions in different clusters.
First, let us understand the differences between the green and
orange clusters. The decision tree shows that the \texttt{max\_feature} parameter
distinguishes these two clusters. While the values `auto' and `None' do not restrict the number
of features during training, `sqrt' and `log2' randomly choose a subset of
features according to the square root
and the base-2 logarithm of total features during training.
This explanation validates findings from Zhang and Harman~\cite{DBLP:conf/icse/ZhangH21}
where they reported that restricting the number of features strengthens
the biases in ML models. Another localized parameter is the minimum sample
to stop the growth and fit leaf models. This distinguishes the orange cluster
from the two other clusters. Intuitively, underprivileged groups
tend to have less representation in the dataset. Since random forests
assign predictions to the majority class in the leaves,
they tend to favor privileged groups when a threshold on the minimum
samples is set.

\noindent\textit{Feature Transferability.} In another experiment, we consider
the $race$ attribute as the protected using the $census$ dataset.
Figure~\ref{fig:overview-TreeRegressor} (c) shows the inputs
generated for the race feature is clustered into two groups.
The explanation tree in Figure~\ref{fig:overview-TreeRegressor} (d)
shows that the max feature and minimum samples in leafs
are two parameters in the tree regressor that explain the difference
in the $AOD$ biases, similar to the case when $sex$ is the protected attribute.

\noindent\textit{Dataset Transferability.} We also study other
datasets in ML fairness literature including the German Credit Data
(Credit)~\cite{Dua:2019-credit} (see Section~\ref{sec:rq3}). For the
random forest, our findings are stable across different datasets and protected attributes: the
minimum sample weights of leaf nodes and maximum features are the most important parameters
to distinguish configurations with high and low biases. These results can help ML developers
understand the fairness implications of different configuration options in their libraries.

\vspace{0.5em}
\noindent\textbf{Mitigation Technique.}
As we discussed previously, \toolname is also useful to suppress biases
by picking low-bias hyperparameters. The details
of experiments to show the mitigation aspect of \toolname can
be found in Section~\ref{sec:rq4}. For random forests, \toolname can
mitigate the biases from an $EOD$ of $11.6\%$ to $0.1\%$ with even better accuracy
compared to the default configuration as a baseline. There are cases that \toolname
alone cannot reduce biases in a statistically significant way. In such cases,
we found that combing \toolname with existing approaches
can significantly reduce biases under certain conditions (see RQ4 in Section~\ref{sec:rq4}).

%% file: overview-Figure.tex
\begin{figure*}[!htb]
    \centering
    \begin{minipage}{0.24\textwidth}
        \centering
   		\includegraphics[width=1\textwidth]{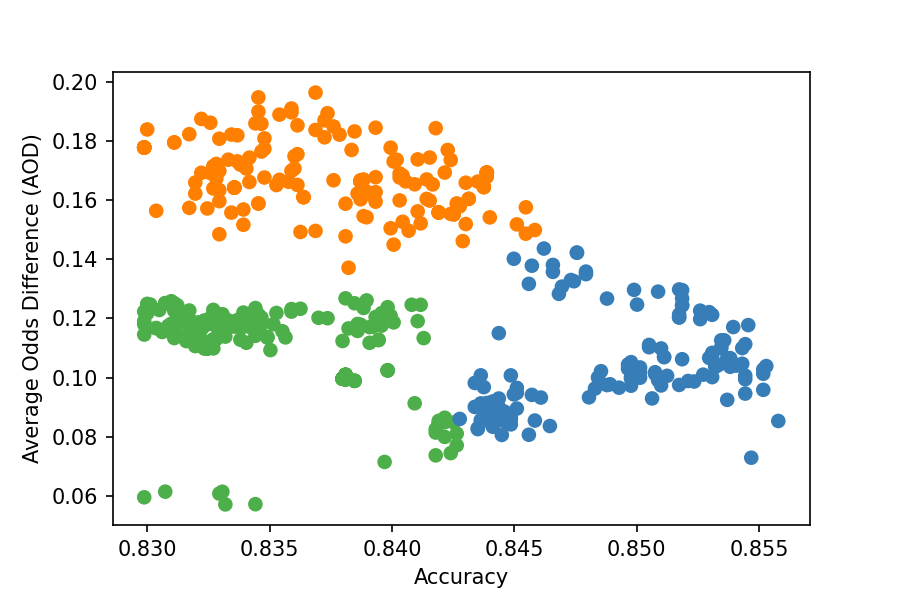}
    \end{minipage}
    \begin{minipage}{0.24\textwidth}
        \centering
        \includegraphics[width=0.8\textwidth]{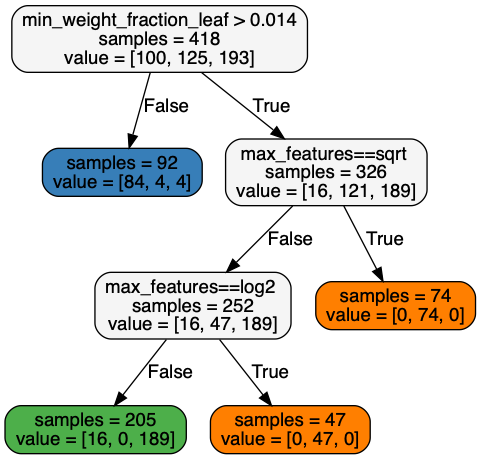}
    \end{minipage}
    \begin{minipage}{0.24\textwidth}
        \centering
   		\includegraphics[width=1\textwidth]{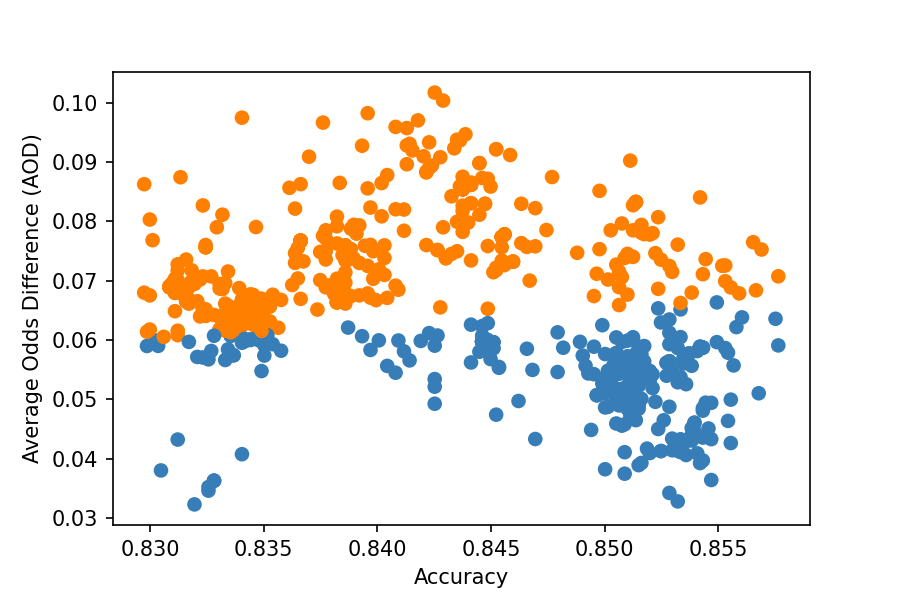}
    \end{minipage}
    \begin{minipage}{0.24\textwidth}
        \centering
        \includegraphics[width=0.85\textwidth]{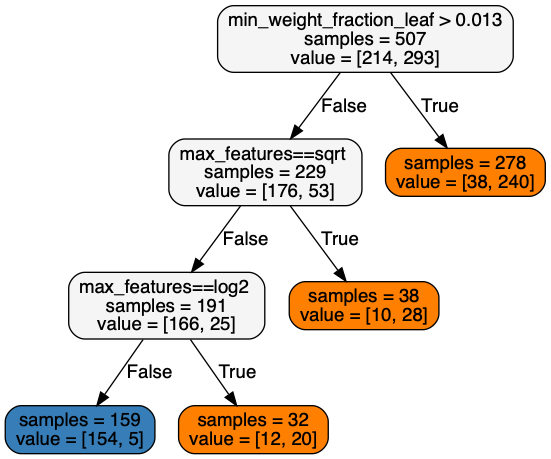}
    \end{minipage}
    \caption{
    (a) Test cases for $census$ with $sex$ are clustered into three:
    y-axis is the $AOD$ bias and x-axis is the accuracy;
    (b) The tree classifier explains that the \texttt{max\_features} and
    the \texttt{min\_weight\_fraction\_leaf} discriminate the three clusters;
    (c) Two clusters for $census$ with $race$;
    (d) The classifier for $census$ with $race$ shows a similar explanation to $census$ with $sex$.}
    \label{fig:overview-TreeRegressor}
\end{figure*}

%% file: definition.tex
The primary performance criteria for data-driven software is
accuracy. However, the presence of fairness results in a multi-objective
optimization problem.
We propose a search-based solution to approximate the curve of Pareto-dominant
hyperparameter configurations
and a statistical learning method to succinctly explain what hyperparameter
distinguish high fairness from low fairness.

\vspace{0.5em}
\noindent{\bf The ML Paradigm.}
Data-driven software systems often deploy mature, off-the-shelf ML libraries to
learn various models from data.
We can abstractly view a learning problem as the problem of identifying a
mapping $M: \Xx \to \Yy$ from a set $\Xx$ of inputs to a set $\Yy$ of outputs by
learning from a fixed dataset $\Dd = \set{({\bf x_i}, {\bf y_i})}_{i=1}^N$ so that
$M$ generalizes well to previously unseen situations.

The application interfaces of these ML libraries expose configuration
parameters---characterizing the set $\Hh$ of hyperparameters---that let the
users define the hypothesis class for the learning tasks.
These hypothesis classes themselves are defined over a set of model parameters
$\Theta_h$ based on the selected hyperparameter $h \in \Hh$.
The ML programs sift through the given dataset $\Dd$ to learn
an ``optimal'' value $\theta \in \Theta_h$ and thus compute the learning model
$M_h(\theta|\Dd): \Xx \to \Yy$ automatically.
When $\Dd$ and $\theta$ are clear from the context, we write $M_h$ for the
resulting model.

The fitness of a hyperparameter $h \in \Hh$ is evaluated by computing the
accuracy (ratio of correct results) of the model $M_h$ on a validation
dataset $\Dd_*$. We denote the accuracy of a model $M$ over $\Dd_*$ as
$ACC^M$.
The dataset $\Dd_*$ is typically distinct from $\Dd$ but assumed to be sampled
from the same distribution.
Hence, the key design challenge for the data-driven software engineering
is a {\it search problem} for optimal configuration of the
hyperparameters maximizing the accuracy over $\Dd_*$.

\vspace{0.5em}
\noindent{\bf Fairness Notion.}
To pose fairness requirements over the learning algorithms, we assume the access
to two predicates.
The predicate $\pi: \Xx \to \set{0, 1}$
over the input variables characterizing the protected status of a data point
${\bf x}$ (e.g., race, sex, or age).
Without loss of generality, we assume there are only two protected groups:
a group with $\pi({\bf x}) = 0$ and a group with $\pi({\bf x}) = 1$.
We also assume that the predicate $\phi: \Yy \to \set{0, 1}$ over the output variables characterizes
a favorable outcome (e.g., low reoffend risk) with $\phi({{\bf y}}) = 1$.

Given $\Dd_*$ and $M: \Xx \to \Yy$, we define
true-positive rate (TPR) and false-positive rate (FPR) for the protect group
$i \in \set{0, 1}$ as
\vspace{0.5em}
\begin{eqnarray*}
  TPR^M(i) &=& \frac{\big|\set{({\bf x}, {\bf y}) \in \Dd_* : \pi({\bf x}) = i,  \phi(M({\bf x})) = 1, \phi({\bf y}) = 1}\big|}
       {\big|\set{({\bf x}, {\bf y}) \in \Dd_* : \pi({\bf x}) = i}\big|}\\
  FPR^M(i) &=& \frac{\big|\set{({\bf x}, {\bf y}) \in \Dd_* : \pi({\bf x}) =
       i, \phi(M({\bf x})) = 1, \phi({\bf y}) = 0}\big|}
       {\big|\set{({\bf x}, {\bf y}) \in \Dd_* : \pi({\bf x}) = i}\big|}.
\end{eqnarray*}
We use two prevalent
notions of fairness~\cite{bellamy2019ai,chakraborty2020fairway,DBLP:conf/icse/ZhangH21}.
Equal opportunity difference (EOD) of $M$ against $\Dd_*$
between two groups is
\vspace{0.1em}
\[
EOD^M = \big|TPR^M(0) - TPR^M(1) \big|,
\]
and average odd difference (AOD) is
\vspace{0.1em}
\[
AOD^M = \frac{|TPR^M(0)-TPR^M(1)| + |FPR^M(0)-FPR^M(1)|}{2}.
\]
Let $\Ff \in \set{EOD^M, AOD^M}$ be some fixed fairness criterion.
We notice that a high value of $\Ff$ implies high bias (low fairness) and a low
value implies low bias (high fairness).

The key design challenge for social-critical data-driven software systems is
to search for fairness- (bias-) optimal configuration $h \in \Hh$ of hyperparameters maximizing the
accuracy and minimizing the bias $\Ff$ of the resulting model $M_h$.
A hyperparameter $h \in \Hh$ is Pareto-fairness-dominated by $g \in \Hh$ if the
model $M_g$ provides better accuracy and lower bias, i.e.
$ACC^{M_h} < ACC^{M_{g}}$ and $\Ff^{M_h} > \Ff^{M_g}$.
We say that a hyperparameter $h \in \Hh$ is Pareto-fairness-optimal if it is not
fairness-dominated by any other hyperparameters.
Similarly, we can define Pareto-bias-domination ($h$ is Pareto-bias-dominated by $g$ if $ACC^{M_h} < ACC^{M_{g}}$ and $\Ff^{M_h} < \Ff^{M_g}$)
 and Pareto-bias-optimal hyperparamters.
A Pareto set is a graphical depiction of all of Pareto-optimal points.
Since we are interested in hyperparameters that lead to either low bias (high
fairness) or high bias (low fairness), our goal is to compute Pareto sets for
both fairness and bias optimal hyperparameters: we call this set a twined Pareto set.
Our goal is to compute a convenient approximation of the twined Pareto set
that can be used to identify, explain,
and exploit the hyperparameter space to improve fairness.

\begin{definition}[hyperparameter Discovery and Debugging]
    \label{def-problem}
    \begin{tcolorbox}[boxrule=1pt,left=1pt,right=1pt,top=1pt,bottom=1pt]
     Given an ML algorithm and a dataset with protected and favorable predicates,
    the \textit{hyperparameter discovery problem} is
    to approximate
    the twined Pareto set (both fairness-optimal and bias-optimal points).
    Given such approximation, the \textit{fairness debugging problem} is
    to explain the difference between the hyperparameter characterizing the high
    and low fairness with acceptable accuracy.
    \end{tcolorbox}
\end{definition}

%% file: approach.tex
We propose dynamic search algorithms
to discover hyperparameters that characterize fairness-optimal and bias-optimal 
models and statistical debugging to localize what hyperparameters
distinguish fair models from biased ones.

\input{algorithm}
\vspace{0.5em}
\noindent \textbf{Hyperparameter Discovery Problem}.
The grid search is an exhaustive method to approximate the Pareto
curve, however, it suffers from the curse of dimensionality.
Randomized search may alleviate this curse to some extent, however, given its
blind nature and the large space of parameters, it may fail to explore
interesting regions.
Evolutionary algorithms (EA) guided by the promising input seeds often
explore extreme regions of the Pareto space and are thus natural candidates for
our search problem. 
While multi-objective EAs look promising,
they are notoriously slow~\cite{8469102}. For example, NSGA-II~\cite{deb2002fast}
has quadratic complexity to pick the next best candidate form the samples in
the population. Instead, we propose a single objective EA
with accuracy constraints.
Algorithm~\ref{alg:overall} sketches our approach for detecting and explaining
strengths of discriminations in the configuration space of ML libraries.

\vspace{0.25em}
\noindent \textit{General Search Algorithms.}
The random search algorithm generates test
inputs uniformly and independently from the domain of parameter variables.
The black-box search is an evolutionary algorithm that
selects and mutates inputs from its population. The gray-box evolutionary search
uses the same strategy as the black-box search,
but is also guided by the code coverage of libraries' internals.

\vspace{0.25em}
\noindent \textit{Initial Seeds.}
Our approach starts with the default configuration and runs the
learning algorithm over the training dataset to build a machine learning model
(line 1 of Algorithm~\ref{alg:overall}).
If the search algorithm is gray-box, the running of algorithm also returns the path
characterizations. A path characterization is \texttt{xor} of hash values
obtained from program line numbers visited in the run. Then, we use the machine
learning model and the validation set to infer the predictions (line 2). We use
the predictions, their ground-truths, and protected attributes to measure
the prediction accuracy and the group fairness metrics such as $EOD$ and $AOD$ (line 3).

\vspace{0.25em}
\noindent \textit{Input Selections.}
We add the default configuration and its outcome to the population (line 4)
and iteratively
search to find configurations that minimize and maximize biases given
a threshold on the accuracy.
In doing so, we consider the type of search
to generate inputs. If the search is ``random'', we randomly and uniformly
sample from the domain of configuration parameters (lines 7 to 8). Otherwise, we pick an
input from the population based on a weighted sampling strategy that prefers more
recent inputs: given a location $i>0$,
the probabilistic weight of sample $i$ is
$\frac{2*i}{n*(n+1)}$ where $n$ is the size of input population, assuming a higher
location is more recent. Then, we randomly
choose a parameter and apply mutation operations over its current value
to generate a new configuration (lines 9 to 11).
We use standard mutation operations such as increasing/decreasing value by a
unit. Given the new configuration, we perform the training and inference
steps to measure the prediction accuracy and biases (lines 12 to 14).

\vspace{0.25em}
\noindent \textit{Search Objective.}
Identifying promising configurations is a critical step in our algorithm.
We consider the characteristics of the new configuration and compare them
to the test corpus (line 15). We say a configuration is promising if
no existing configuration in the test corpus Pareto-fairness-dominate or
Pareto-bias-dominate (based on $EOD$ and $AOD$) the new configuration.
Thus, we add promising inputs to the test corpus (line 16).
If the search type is gray-box, we also consider the path characterization.
If the path has not been visited before and the corresponding configuration
manifests an accuracy equal to or better than
the accuracy of the default configuration within $\epsilon=1.0\%$ margin,
we add the configuration to the population as well.

\vspace{0.5em}
\noindent \textbf{Fairness Debugging Problem.}
With the assumption that the hyperparameter space $\Hh$ is given as a finite set
of hyperparameter variables $\Hh_1 \times  \Hh_2 \times  \ldots \times \Hh_m$,
we wish to explain
the dependence of these individual hyperparameter variables towards fairness for
all Pareto-optimal hyperparameters as approximated in the test corpus $I$.
Our explanatory approach uses clustering in the domain of fairness
vs. accuracy to discover $k$ classes of hyperparameter
configurations in the test corpus $I$ (line 17). Then, we use standard decision tree
classifiers to generate succinct and interpretable predicates over the
hyperparameter variables (line 18).
The resulting $k$ predicates serve as an explanatory
model to understand biases in the configuration of learning algorithms.

%% file: algorithm.tex
\begin{algorithm}[t!]
{
	\DontPrintSemicolon
	\KwIn{algorithm $\Aa$, space of hyperparamters $\Hh$, default configuration $h_{d}$,
		training dataset ($X_T,y_T$),
		test dataset ($X_t,y_t$), protected attribute $A$,
		type of search $S_t$, the margin $\epsilon$,
		time-out $T$, num. clusters $k$.
	}
	\KwOut{Test Cases $I$, Predicates $\Phi$.}

	$model$, $path$ $\gets$ \texttt{run}($\Aa$, $h_{d}$, $X_T$, $y_T$, $S_t$)

	$pred$ $\gets$ \texttt{infer}($model$, $X_t$)

	$accuracy_d$, $fairness_d$ $\gets$ \texttt{metric}($pred$, $y_t$, $A$)

	$I$.\texttt{add}($h_d$, $accuracy_d$, $fairness_d$, $path$)

	$cur$ $\gets$ \texttt{time}()

	\While{\texttt{time}() - $cur$ $<$ $T$}{

		\If{$S_t$ $==$ ``random''}{
			$h$ $\gets$ \texttt{UniformlyRandom}($\Hh$)
		}\ElseIf{$S_t$ $==$ ``black-box'' \texttt{or} $S_t$ $==$ ``gray-box'' }{
			$h'$ $\gets$ \texttt{choice$_{W}$}($I$.\texttt{project}($\Hh$))

			$h$ $\gets$ \texttt{mutate}($h'$)
		}

		$model$, $path$ $\gets$ \texttt{run}($\Aa$, $h$, $X_T$, $y_T$, $S_t$)

		$pred$ $\gets$ \texttt{infer}($model$, $X_t$)

		$accuracy, fairness$ $\gets$ \texttt{metric}($pred$, $y_t$, $A$)

		\If{\texttt{promising}($h$, $accuracy$, $\epsilon$, $fairness$, $path$, $I$)}{
				$I$.\texttt{add}($h$, $accuracy$, $fairness$, $path$)
		}
	}

	$label$ $\gets$ \texttt{spectralClust}($I.$\texttt{project}($accuracy$, $fairness$), $k$)

	$\Phi$ $\gets$ \texttt{DTClassifier}($I.$\texttt{project}($\Hh$), $label$)


	\caption{\textsc{\toolname: Detecting and Explaining Fairness and Bias
	in parameters of ML algorithms.}}
	\label{alg:overall}
}
\end{algorithm}

%% file: experiments.tex
We first pose research questions. Then, we elaborate
on the case studies, datasets, protected attributes,
our tool, and environment. Finally, we
carefully examine and answer research questions.

\begin{enumerate}[start=1,label={\bfseries RQ\arabic*},leftmargin=3em]

\item What is the magnitude of biases in the \textit{hyperparameter} space of
ML algorithms?

\item Are mutation-based and code coverage-based evolutionary algorithms
effective to find interesting configurations?

\item Is statistical debugging useful to explain the biases in the
\textit{hyperparameter} space of ML algorithms? Are these parameters consistent
across different fairness applications?

\item Is our approach effective to mitigate biases as compared
to the state-of-the-art \textit{inprocess} technique?

\end{enumerate}
All subjects, experimental results, and our tool are available
on our GitHub repository: \url{https://github.com/Tizpaz/Parfait-ML}.

{\footnotesize
\begin{table*}[!t]
\caption{Datasets used in our experiments.}
\centering
\resizebox{0.8\textwidth}{!}{
\begin{tabu}{|l|l|l|ll|ll|}
  \hline
  \multirow{2}{*}{\textbf{Dataset}} & \multirow{2}{*}{\textbf{|Instances|}} & \multirow{2}{*}{\textbf{|Features|}} & \multicolumn{2}{c|}{\textbf{Protected Groups}} & \multicolumn{2}{c|}{\textbf{Outcome Label}} \\
   &  & & \textit{Group1} & \textit{Group2} & \textit{Label 1} & \textit{Label 0}  \\
  \hline
  Adult \textit{Census} & \multirow{2}{*}{$48,842$} & \multirow{2}{*}{$14$} & Sex-Male & Sex-Female & \multirow{2}{*}{High Income} & \multirow{2}{*}{Low Income} \\ \cline{4-5}
  Income &  &    &  Race-White & Race-Non White  &   &    \\
  \hline
  \multirow{2}{*}{\textit{Compas} Software}  & \multirow{2}{*}{$7,214$} & \multirow{2}{*}{$28$} & Sex-Male & Sex-Female & \multirow{2}{*}{Did not Reoffend} & \multirow{2}{*}{Reoffend} \\  \cline{4-5}
  &  &    &  Race-Caucasian & Race-Non Caucasian  &   &    \\
  \hline
  German \textit{Credit}  & $1,000$ & $20$ & Sex-Male & Sex-Female & Good Credit & Bad Credit \\
  \hline
  \textit{Bank} Marketing  & $45,211$ & $17$ & Age-Young & Age-Old & Subscriber & Non-subscriber \\
  \hline
\end{tabu}
}
\label{table:dataset}
\end{table*}
}

\subsection{Subjects}
\label{sec:subjects}
We consider $5$ ML algorithms
from the literature~\cite{FairnessTesting,udeshi2018automated,10.1145/3338906.3338937,chakraborty2020fairway}:

\vspace{0.25em}
\noindent \textit{1) Logistic regression (LR)} uses sigmoid functions to map input data to a
real-value outcome between 0 and 1. We use an implementation from scikit-learn~\cite{logistic-regression} that has $15$ parameters including three booleans, three categoricals,
four integers, four reals, and one dictionary.
Example parameters are the norm of penalization, prime vs dual formulation, and
tolerance of optimization.

\vspace{0.25em}
\noindent \textit{2) Random forest (RF)} is an ensemble method that fits a number
of decision trees and uses the averaged outcomes for predictions.
We refer to Overview Section~\ref{sec:overview} for further information.

\vspace{0.25em}
\noindent \textit{3) Support vector machine (SVM)} is a classifier that finds
hyperplanes to separate classes and maximizes margins between
them. The scikit-learn implementation has $12$ parameters including
two booleans, three categoricals, three integers, three reals,
and one dictionary~\cite{SVM}. Examples are tolerance and regularization
term.

\vspace{0.25em}
\noindent \textit{4) Decision tree (DT)} learns decision logic from input data in the
form of if-then-else statements. We use an implementation that has $13$ parameters
including three categoricals, three integers, six reals, and one dictionary~\cite{Decision-Tree}.
Example parameters are the minimum samples in the node to split
and maximum number of leaf nodes.

\vspace{0.25em}
\noindent \textit{5) Discriminant analysis (DA)} fits data to a Gaussian prior of class labels.
Then, it uses the posterior distributions to predict the class of new data.
We use an implementation that has $11$ parameters including two booleans, one categorical,
one integer, four reals, two lists, and one function~\cite{Discriminant-Analysis}.

We also consider four datasets with different protected attributes and define
six training tasks as shown in Table~\ref{table:dataset}, similar to prior
work~\cite{chakraborty2020fairway,FairnessTesting,10.1145/3338906.3338937}.
Adult Census Income ($census$)~\cite{Dua:2019-census},
German Credit Data ($credit$)~\cite{Dua:2019-credit},
Bank Marketing ($bank$)~\cite{Dua:2019-bank}, and COMPAS Software ($compas$)~\cite{compas-dataset}
are binary classification tasks to predict whether an individual has income over $50$K,
has a good credit history, is likely to subscribe, and has a low reoffending risk,
respectively.

\subsection{Technical Details}
\label{subsec:tech-detailts}
Our tool has detection and explanation components. The detection component
is equipped with three search algoirthms: \textit{random}, \textit{black-box} mutations,
and \textit{gray-box} coverage. The search algorithms are described
in Approach Section~\ref{sec:approach}.
We implement these techniques in Python where
we use the XML parser library to define the parameter variables and their domains
and Trace library~\cite{Trace} to instrument programs for the code coverage~\cite{fuzzingbook2021}.
We implement the clustering using Spectral algorithm~\cite{von2007tutorial}
and the tree classifier using the CART algorithm~\cite{Breiman/1984/CART} in
scikit-learn~\cite{Decision-Tree}.

\subsection{Experimental Setup}
\label{subsec:exper-setup}
We run all the experiments on a super-computing machine with the Linux Red Hat $7$ OS and an
Intel Haswell $2.5$ GHz CPU  with 24 cores (each with $128$ GB of RAM).
We use Python $3.6$ and scikit-learn version $0.22.1$.
We set the timeout for our search algorithm to $4$ hours
for all experiments unless otherwise specified.
Additionally, each experiment has been repeated $10$
times to account for the randomness of search techniques. We averaged
the results and calculated the $95$\% confidence intervals to report results.
The difference between two means is statistically
significant if their confidence intervals do not overlap~\cite{https://doi.org/10.1002/stvr.1486}.
We split the dataset into training data ($75\%$) and validation data ($25\%$).
We train an ML model with a given learning algorithm, its configuration,
and the training data. Finally, we report the accuracy and fairness metrics
over the inferred ML model using the validation set.
Any configurations that achieve higher accuracy than the default
configuration (with $1\%$ margins) are \textit{valid inputs}.

\subsection{Magnitude of Biases (RQ1)}
\label{sec:rq1}
\input{RQ-1-Table}

One crucial research question in this paper is to understand the magnitude
of biases when tuning hyperparameters.
Table~\ref{table:bias-magnitude} shows the magnitude of biases observed
for different learning algorithms over a specific dataset and protected attribute.
We consider the inputs from all search algorithms and report the average
as well as $95$\% confidence intervals (in the parenthesis) of different metrics.
The column $Num. Inputs$ shows the number of valid test cases generated from the
detection step. The column $Accuracy_{range}$ shows
the range of accuracies observed from all generated configurations.
The column $AOD_{range}$ shows the
range of $AOD$ biases for all configurations;
$AOD_{min}^{top}$ shows the lowest $AOD$ biases for inputs within top $1\%$ of
prediction accuracy; $AOD_{max}^{top}$ shows
the highest biases for inputs within the top $1\%$ of accuracy.
For the example of $DT$ with $census$ and $race$, $AOD_{range}$ shows
the $AOD$ biases for configurations within
$79.2$\% to $84.7$\% accuracy, whereas $AOD_{min}^{top}$ shows the
lowest biases within $83.7\%$ to $84.7\%$ accuracy.
The column $EOD_{range}$, $EOD_{min}^{top}$,
and $EOD_{max}^{top}$ show the range of biases based on equal opportunity difference
($EOD$) for all valid inputs, the lowest $EOD$ biases for inputs within the top $1\%$ of
accuracy, and the highest biases for inputs
within top $1$\% of accuracy.

The results
show that the configuration of hyperparameters indeed amplifies and suppresses ML biases.
Within $1$\% of
(top) accuracy margins, a fairness-aware configuration can suppress the group biases
to below $1\%$ for $AOD/EOD$, and a poor choice can amplify the biases up to $23\%$
for $EOD$ and up to $15\%$ for $AOD$.

\begin{tcolorbox}[boxrule=1pt,left=1pt,right=1pt,top=1pt,bottom=1pt]
\textbf{Answer RQ1:}
Tuning of hyperparameters significantly
affects fairness. Within $1$\% of accuracy margins,
a fairness-aware configuration can reduce the EOD bias to below $1$\% and a poor choice
of configuration can amplify the EOD bias to $23$\%.
\end{tcolorbox}

\subsection{Search Algorithms (RQ2)}
\label{sec:rq2}
\input{RQ-2-Table}
In this section, we compare the results of three search algorithms
to understand which method is more effective in finding configurations
with low and high biases. Table~\ref{table:search-bias} shows the
number of generated valid inputs per search method, the absolute
difference between the maximum $AOD$ and the minimum $AOD$, and
the absolute difference between the maximum $EOD$ and the minimum $EOD$.
The results show that there are multiple statistically significant
difference among the three search strategies. In $4$ cases,
the random strategy generates
the lowest number of inputs.
In $5$ cases, the evolutionary algorithms (both \textit{black-box} and \textit{gray-box})
outperforms the random stratgey in finding configurations that characterize
significant $EOD$ and $AOD$ biases.

The comparison between black-box and gray-box evolutionary algorithms shows that
there is no statistically significant
difference between them in generating configurations
that lead to the lowest and highest biases. We conjecture
that code coverage in detecting biases is not particularly useful since the
biases are not introduced as a result of mistakes in the code implementation, rather they are
results of unintentionally \textit{choosing} poor configurations of learning algorithms
by ML users or \textit{allowing} poor configurations of algorithms
by ML library developers in fairness-sensitive applications.

\input{RQ-2-Figures}

In Table~\ref{table:search-bias}, we observe that the
statistically significant differences are relevant to the
decision tree (DT). For the algorithm, we provide the temporal
progress of three search algorithms for different training scenarios
(see supplementary material for the rest).
Figure~\ref{fig:search-plots} shows the mean of maximum biases (sold line)
and the $95\%$ confidence intervals (filled colors) over the $4$ hours testing
campaigns of each search strategy. There is a statistically significant
difference if white spaces are present between the confidence intervals.

\begin{tcolorbox}[boxrule=1pt,left=1pt,right=1pt,top=1pt,bottom=1pt]
\textbf{Answer RQ2:}
Our experiments show that mutation-based evolutionary algorithms
are more effective in generating configurations that characterize low
and high bias configurations.
We did not find a statistically significant
difference to support using code coverage
in fairness testing of learning libraries.
\end{tcolorbox}

\subsection{Statistical Learning for Explanations (RQ3)}
\label{sec:rq3}
We present a statistical learning approach to explain what
configurations distinguish low bias models from high bias ones. We
use clustering to find different classes of biases
and the CART tree classifiers to synthesize predicate functions
that explain what parameters are common in the same cluster
and what parameters distinguish one cluster from another.
Similar techniques have been used for software performance
debugging~\cite{DPFUZZ}.
We limit the maximum number of clusters to $3$ and prefer $three$
clusters over $two$ clusters if and only if the corresponding classifier
achieves better accuracy. We also limit the depth of CART classifiers to $3$
in order to generate succinct decision trees.
We first present the explanatory
models of different algorithms over each individual training scenario
(e.g, $census$ dataset with $sex$).
Next, we perform an aggregated analysis of learning
algorithms over the $6$ different training datasets, the $3$ random search
algorithms, and the $10$ repeated runs to extract what hyperparameters
are frequently appearing in the explanatory models and
thus suspicious of influencing fairness systematically.

\input{RQ-3-Figures}

\vspace{0.5em}
\noindent \textbf{Individual training scenario}.
We show how our statistical learning approach helps localize
hyperparameters that influence the biases for each individual training task.
Figure~\ref{fig:clustering-explanation} shows the explanatory models
(clustering and CART tree) for each learning algorithm over the $census$
dataset with $sex$ (except for random
forest that was presented in the overview Section~\ref{sec:overview}).
For example,
Figure~\ref{fig:clustering-explanation} (b)
shows the true evaluation of ``\texttt{solver!=sag}
$\land$ \texttt{fit-intercept>0.5} $\land$ \texttt{solver=newton-cg}''
for the hyperparameters of logistic regression,
which explains the orange cluster, leads to stronger biases.
All models are available in the supplementary material.

\vspace{0.5em}
\noindent \textbf{Mining over all training scenarios}.  For each
learning algorithm, our goal is to understand what hyperparameters
influence the fairness in multiple training scenarios and establish
whether some hyperparameters systematically influence fairness beyond
a specific training task. Overall, we analyze $180$ CART trees
and report hyperparameters that appear as a node
in the tree more than $50$ times overall
and more than $3$ times uniquely in the $4$ datasets.
Different values of these frequent hyperparameters are suspicious of
introducing biases, across datasets, search algorithms, and
different protected attributes.  In the following, within the
parenthesis right after the name of a hyperparameter, we report (1)
the number of explanatory models (out of 180) where the hyperparameter
appears and (2) the number of datasets (out of 4) for which there
is an experiment whose explanatory model contains the hyperparameter.

\vspace{0.25em}
\noindent \textit{A) Logistic regression (LR)}:
The computation time for inferring clusters and tree classifiers
is $77.9$ (s) in the worst case.
The accuracy of classifiers is between $90.9\%$ and $96.8\%$.
Three frequent hyperparameters based on
the predicates in classifiers are \texttt{solver} (175,4), \texttt{tol} (53,3),
and \texttt{fit-intercept} (50,3). Our analysis shows that
the solver $saga$ frequently achieves low biases after tuning the tolerance
parameter whereas the solver $newton$-$cg$ often achieves low biases
if the intercept term is added to the decision function.

\vspace{0.25em}
\noindent \textit{B) Random forest (RF)}:
The computation time for
inferring models is $75.7$ (s) in the worst case.
The accuracy is between $80.5\%$ and $100.0\%$.
Two frequent parameters are \texttt{max\_features} (170,4)
and \texttt{min\_weight\_fraction\_leaf} (160,4). These parameters
and their connections to fairness are explained in the overview section~\ref{sec:overview}.

\vspace{0.25em}
\noindent \textit{C) Support vector machine (SVM)}:
The computation time for
inferring models is $79.9$ (s) in the worst case.
The accuracy is between $83.9\%$ and $98.3\%$.
The only (relatively) frequent parameter is \texttt{degree} (53,3). This shows the high variation of
parameter appearances in the explanation model. Thus, the configuration of SVM
might not systematically amplify or suppress biases;
the influence of configuration on biases largely depends on the specific training task.

\vspace{0.25em}
\noindent \textit{D) Decision tree (DT)}:
The computation time for inferring models
is $76.9$ (s) in the worst case.
The accuracy is between $93.8\%$ and $98.1\%$.
The frequent parameters
are
\texttt{min\_fraction\_leaf} (114,4) and
\texttt{max\_features} (114, 4). Similar to random forest,
the minimum required samples in the leaves and
the search space of dataset attributes during training impact fairness
systematically.

\vspace{0.25em}
\noindent \textit{E) Discriminant analysis (DA)}:
The computation time for
inferring models is $77.8$ (s) in the worst case.
The accuracy is between $54.2\%$ and $93.8\%$. However, if
we allowed a higher depth for the classifier (more than $3$), it is
above $90\%$ in all cases. The frequent parameter is
\texttt{tol} (141,4). However, the exact condition over
the tolerance in the explanatory model
significantly depends on the training
task, and might not influence fairness systematically.

ML library maintainers can use these results to understand the fairness implications
of their library configurations.

\vspace{0.5em}
\vspace{0.5em}
\begin{tcolorbox}[boxrule=1pt,left=1pt,right=1pt,top=1pt,bottom=1pt]
\textbf{Answer RQ3:}
We found the statistical learning scalable and useful to explain and
distinguish the configuration with low and high biases.
Our global analysis of $180$ explanatory models
per learning algorithm reveals that some algorithms and their configurations
can systematically amplify or suppress biases.
\end{tcolorbox}

\subsection{Bias Mitigation Algorithms (RQ4)}
\label{sec:rq4}
\input{RQ-4-Table}

We show how \toolname can be used as a mitigation tool
to aid ML users pick a configuration of algorithms with the lowest discrimination.
In doing so, we run \toolname for a short amount of time and pick a
configuration of hyperparameters with the lowest $AOD$ and $EOD$.
To show the effectiveness, we compare \toolname to the state-of-the-art
techniques~\cite{agarwal2018reductions,chakraborty2019software,chakraborty2020fairway}.
We say approach (1) outperforms approach (2) if it achieves statistically significant
lower biases within a similar or higher accuracy.

\vspace{0.5em}
\noindent \textit{A) Exponentiated gradient}~\cite{agarwal2018reductions} presents a bias reduction
technique that maximizes the prediction accuracy subject to linear
constraints on the group
fairness requirements. They use Lagrange methods and apply the exponentiated gradient
search to find Lagrange multipliers to balance accuracy and fairness. In doing
so, they use meta-learning algorithms to learn a family of
classifiers (one in each step of the algorithm with
a fixed Lagrange multiplier) and assign a probabilistic weight to each of them.
In the prediction stage, the approach chooses
one classifier from the family of classifiers stochastically according
to their weights.
We choose this approach for a few reasons: 1) the approach is an \textit{inprocess}
algorithm and does not add or remove input data samples in the pre-processing step
nor modifies prediction labels in the post-processing step;
2) they assume black-box access to ML algorithms;
thus they do not modify the learning objective nor model parameters.
We note that their approach is sensitive to the fairness metric (to construct
the linear constraints) and does not support arbitrary metrics. In particular,
they support the $EOD$ metric, but not the $AOD$ metric. Thus, we focus on the
$EOD$ metric in this experiment.
In addition, since the discriminant analysis algorithm does not support
meta-learning, we exclude this algorithm from this experiment.

We consider the default configuration of algorithms without any fairness
consideration, the exponentiated gradient method~\cite{agarwal2018reductions},
\toolname, and exponentiated gradient combined with \toolname. We also
set the execution time of \toolname to $6$ minutes in accordance with the (max)
execution time of gradient approach in our environment.
Table~\ref{table:comparison} shows the results of these
experiments. Compared to the default configuration, in $18$ cases out of $24$ experiments,
the exponentiated gradient significantly reduces the $EOD$ biases. However,
in $11$ cases out of $24$ experiments, exponentiated gradient degraded
the prediction accuracy. \toolname reduces the $EOD$ biases in $23$ cases with
$12$ cases of accuracy improvements and only one case of accuracy degradations.
Overall, \toolname significantly outperforms
the gradient method (discrepancies are highlighted with red font in Table~\ref{table:comparison}).
Combining \toolname
and exponentiated gradient performs better than each technique in isolation
(see $RF$ with $census$ and $sex$) given that the gradient technique does not
increase the strength of biases in isolation (see $LR$ with $credit$ and $sex$).
In such cases, \toolname alone results in lower biases and higher accuracy.

\input{RQ-4-Table-2}

\vspace{0.5em}
\noindent \textit{B) }\textsc{Fairway}~\cite{chakraborty2019software,chakraborty2020fairway}
uses a mutli-objective optimization technique known as
\textsc{FLASH}~\cite{8469102} to tune hyperparameters
and chooses a configuration that achieves less biases with a minimum accuracy loss.
We use their implementatio~\cite{Fairway-tool}, and compare our search-based technique to
this method. \textsc{Fairway} generally supports
integer and boolean hyperparameters. Therefore, we use a subset of configurations as
specified and reported for logistic regression (LR)~\cite{chakraborty2020fairway}
and decision tree (DT)~\cite{chakraborty2019software}.
For a fair comparison, we calculate the execution time of \textsc{Fairway} for each
experiment in our environment and limit the execution time of \toolname accordingly.
Table~\ref{table:comparison-FLASH} shows the comparison results. Overall,
there are $3$ discrepancies in AOD and $5$ discrepancies in EOD (noted by red font
in Table~\ref{table:comparison-FLASH}). \toolname outperforms
\textsc{Fairway} in $6$ cases whereas \textsc{Fairway} outperforms \toolname in $2$ cases.
We also noted that the current implementations of \textsc{FLASH} carefully
selected $4$ hyperparameters for LR and DT. When we include three more
hyperparameters with integer or boolean types (e.g., dual and fit\_intercept in $LR$),
we observe that the performance of \textsc{Fairway} significantly degraded. For
$LR$ algorithm over $compas$ with $sex$ scenario, the prediction accuracy is decreased
to $89.7\% (+/- 6.0\%)$, while the AOD and EOD bias metrics are increased to
$2.9\% (+/- 1.0\%)$ and	$3.1\% (+/- 2.3\%)$, respectively. Since \textsc{Fairway}
is sensitive to the domain of variables, \toolname can complement it with the
explanatory model to carefully choose hyperparameters to include in the \textsc{Fairway} search.

\vspace{0.5em}
\begin{tcolorbox}[boxrule=1pt,left=1pt,right=1pt,top=1pt,bottom=1pt]
\textbf{Answer RQ4:}
\toolname is effective to improve fairness by finding low-bias configurations
of hyperparameters. It outperforms exponentiated gradient~\cite{agarwal2018reductions} and
\textsc{Fairway}~\cite{chakraborty2020fairway,chakraborty2019software}
in reducing AOD and EOD biases with equal or better accuracy.
\toolname can complement both approaches to improve fairness.
\end{tcolorbox}

%% file: RQ-1-Table.tex
{\footnotesize
\begin{table*}[ht]
\caption{The magnitude of biases in the parameters of ML algorithms based on
AOD and EOD.}
\centering
\resizebox{\textwidth}{!}{%
\begin{tabu}{|l|l|l|l|l|lll|lll|}
\hline
\multirow{2}{*}{\textbf{Algorithm}} & \multirow{2}{*}{\textbf{Dataset}} & \multirow{2}{*}{\textbf{Protected}} & \multirow{2}{*}{\textbf{Num. Inputs}} & \multirow{2}{*}{\textbf{Accuracy}$_{range}$} & \multicolumn{3}{c|}{\textbf{Average Odds Difference (AOD)}} & \multicolumn{3}{c|}{\textbf{Equal Opportunity Difference (EOD)}} \\
 &  &  &  &  & \textit{AOD}$_{range}$ & \textit{AOD}$_{min}^{top}$ & \textit{AOD}$_{max}^{top}$ & \textit{EOD}$_{range}$ & \textit{EOD}$_{min}^{top}$ & \textit{EOD}$_{max}^{top}$ \\
\hline


\multirow{6}{*}{LR} 
 & Census & Sex & 10,368 (+/- 3,040) & 79.6\% (+/- 0.0\%)-81.1\% (+/- 0.0\%) & 0.3\% (+/- 0.0)-12.4\% (+/- 0.6\%) & 0.7\% (+/- 0.0\%) & 12.0\% (+/- 0.2\%) & 0.1\% (+/- 0.0\%)-23.0\% (+/- 0.0\%) & 0.1\% (+/- 0.1\%) & 23.0\% (+/- 0.0\%) \\
 & Census & Race & 7,146 (+/- 1,699) & 79.7\% (+/- 0.0\%)-81.1\% (+/- 0.0\%) & 0.5\% (+/- 0.2\%)-15.1\% (+/- 1.3\%) & 1.4\% (+/- 0.2\%) & 11.4\% (+/- 0.0\%) & 0.3\% (+/- 0.2\%)-21.0\% (+/- 2.3\%) & 1.5\% (+/- 0.2\%) & 15.7\% (+/- 0.1\%)  \\
 & Credit & Sex & 28,180 (+/- 9,887) & 73.6\% (+/- 0.0\%)-77.2\% (+/- 0.0\%) & 0.9\% (+/- 0.0\%)-13.2\% (+/- 0.4\%) & 1.8\% (+/- 0.2\%) & 8.3\% (+/- 0.7\%) & 0.3\% (+/- 0.1\%)-24.6\% (+/- 0.9\%) & 1.5\% (+/- 0.7\%) & 14.5\% (+/- 1.7\%)  \\
 & Bank & Age & 2,381 (+/- 400) & 88.1\% (+/- 0.0\%)-89.6\% (+/- 0.0\%) & 0.1\% (+/- 0.0\%)-8.9\% (+/- 0.1\%) & 0.1\% (+/- 0.0\%) & 6.7\% (+/- 0.0\%) & 0.0\% (+/- 0.0\%)-15.0\% (+/- 0.1\%) & 0.0\% (+/- 0.0\%) & 12.3\% (+/- 0.0\%)  \\
 & Compas & Sex & 67,736 (+/- 1,832) & 96.0\% (+/- 0.0\%)-97.1\% (+/- 0.0\%) & 1.6\% (+/- 0.0\%)-5.3\% (+/- 0.2\%) & 1.6\% (+/- 0.0\%) & 5.0\% (+/- 0.2\%) & 0.0\% (+/- 0.0\%)-6.2\% (+/- 0.5\%) & 0.0\% (+/- 0.0\%) & 5.9\% (+/- 0.5\%) \\
 & Compas & Race & 66,228 (+/- 3,169) & 96.0\% (+/- 0.0\%)-97.1\% (+/- 0.0\%) & 1.4\% (+/- 0.0\%)- 4.2\% (+/- 0.1\%) &	1.4\% (+/- 0.0\%) & 4.2\% (+/- 0.1\%) & 0.0\% (+/- 0.0\%)-5.1\% (+/- 0.2\%) & 0.0\% (+/- 0.0\%) & 5.1\% (+/- 0.2\%) \\
\hline
\multirow{6}{*}{RF} 
 & Census & Sex & 620 (+/- 105) & 83.0\% (+/- 0.0\%)-85.7\% (+/- 0.0\%) & 5.5\% (+/- 0.6\%)-18.9\% (+/- 0.1\%) & 7.0\% (+/- 0.3\%) & 14.6\% (+/- 0.3\%) & 4.8\% (+/- 0.4\%)-32.3\% (+/- 0.3\%) & 7.5\% (+/- 0.7\%) & 23.0\% (+/- 0.7\%)  \\
 & Census & Race & 605 (+/- 122) & 83.0\% (+/- 0.0\%)-85.7\% (+/- 0.0\%) & 3.2\% (+/- 0.2\%)-10.1\% (+/- 0.3\%) & 3.6\% (+/- 0.1\%) & 9.4\% (+/- 0.2\%) & 4.5\% (+/- 0.3\%)-17.3\% (+/- 0.5\%) & 4.8\% (+/- 0.2\%) & 15.5\% (+/- 0.3\%) \\
 & Credit & Sex & 24,213 (+/- 10,274) & 73.2\% (+/- 0.0\%)-79.2\% (+/- 0.2\%) & 0.1\% (+/- 0.0\%)-15.1\% (+/- 0.2\%) & 2.5\% (+/- 0.4\%) & 6.8\% (+/- 1.2\%) & 0.0\% (+/- 0.0\%)-24.3\% (+/- 0.7\%) & 1.9\% (+/- 1.4\%) & 9.8\% (+/- 2.0\%) \\
 & Bank & Age & 348 (+/- 66) & 89.0\% (+/- 0.0\%)-90.2\% (+/- 0.0\%) & 0.1\% (+/- 0.0\%)-3.1\% (+/- 0.1\%) & 0.0\% (+/- 0.0\%) & 3.0\% (+/- 0.0\%) & 0.0\% (+/- 0.0\%)-5.5\% (+/- 0.3\%) & 0.0\% (+/- 0.0\%) & 5.3\% (+/- 0.3\%)  \\
 & Compas & Sex & 23,975 (+/- 2,931) & 95.5\% (+/- 0.0\%)-97.1\% (+/- 0.0\%) & 1.5\% (+/- 0.0\%)-5.1\% (+/- 0.2\%) & 1.5\% (+/- 0.0\%) & 4.5\% (+/- 0.2\%) & 0.0\% (+/- 0.0\%)-7.1\% (+/- 0.3\%) & 0.0\% (+/- 0.0\%) & 5.8\% (+/- 0.4\%) \\
 & Compas & Race & 22,626 (+/- 3,105) & 95.5\% (+/- 0.0\%)-97.1\% (+/- 0.0\%) & 1.5\% (+/- 0.0\%)-4.6\% (+/- 0.2\%) & 1.5\% (+/- 0.0\%) & 3.7\% (+/- 0.2\%) & 0.0\% (+/- 0.0\%)-6.4\% (+/- 0.3\%) & 0.0\% (+/- 0.0\%) & 4.5\% (+/- 0.3\%) \\
\hline
\multirow{6}{*}{SVM} 
 & Census & Sex & 5,573 (+/- 496) & 65.6\% (+/- 0.1\%)-81.3\% (+/- 0.0\%) & 0.0\% (+/- 0.0\%)-32.6\% (+/- 0.1\%) & 0.2\% (+/- 0.1\%) & 13.3\% (+/- 0.8\%) & 0.0\% (+/- 0.0\%)-29.5\% (+/- 0.9\%) & 0.0\% (+/- 0.0\%) & 17.7\% (+/- 1.1\%) \\
 & Census & Race & 4,595 (+/- 583) & 65.6\% (+/- 0.1\%)-81.3\% (+/- 0.0\%) & 0.0\% (+/- 0.0\%)-30.6\% (+/- 0.8\%) & 0.4\% (+/- 0.0\%) & 12.4\% (+/- 0.8\%) & 0.0\% (+/- 0.0\%)-37.8\% (+/- 1.3\%) & 0.1\% (+/- 0.0\%) & 17.1\% (+/- 1.1\%) \\
 & Credit & Sex & 96,226 (+/- 1,042) & 59.3\% (+/- 5.7\%)-76.5\% (+/- 0.1\%) & 0.0\% (+/- 0.0\%)-17.5\% (+/- 0.0\%) & 1.8\% (+/- 0.4\%) & 9.4\% (+/- 0.4\%) & 0.0\% (+/- 0.0\%)-24.8\% (+/- 0.5\%) & 1.5\% (+/- 0.9\%) & 16.3\% (+/- 0.8\%) \\
 & Bank & Age & 1,361 (+/- 163) & 88.6\% (+/- 0.0\%)-89.8\% (+/- 0.0\%) & 0.0\% (+/- 0.0\%)-5.3\% (+/- 0.4\%) & 0.0\% (+/- 0.0\%) & 5.3\% (+/- 0.4\%) & 0.0\% (+/- 0.0\%)-9.2\% (+/- 0.5\%) & 0.0\% (+/- 0.0\%) & 9.2\% (+/- 0.5\%) \\
 & Compas & Sex & 40,287 (+/- 417) & 96.1\% (+/- 0.0\%)-97.1\% (+/- 0.0\%) & 1.6\% (+/- 0.0\%)-3.8\% (+/- 0.1\%) & 1.6\% (+/- 0.0\%) & 3.8\% (+/- 0.1\%) & 0.0\% (+/- 0.0\%)-3.9\% (+/- 0.2\%) & 0.0\% (+/- 0.0\%) & 	3.9\% (+/- 0.2\%) \\
 & Compas & Race & 40,391 (+/- 540) & 96.1\% (+/- 0.0\%)-97.1\% (+/- 0.0\%) & 1.4\% (+/- 0.0\%)-3.0\% (+/- 0.0\%) & 1.4\% (+/- 0.0\%) & 2.9\% (+/- 0.0\%) & 0.0\% (+/- 0.0\%)-2.9\% (+/- 0.1\%) & 0.0\% (+/- 0.0\%) & 2.8\% (+/- 0.1\%) \\
\hline
\multirow{6}{*}{DT} 
 & Census & Sex & 4,949 (+/- 1,288) & 79.2\% (+/- 0.0\%)-84.9\% (+/- 0.2\%) & 0.3\% (+/- 0.0\%)-32.1\% (+/- 1.8\%) & 5.8\% (+/- 0.6\%) & 13.2\% (+/- 1.4\%) & 0.2\% (+/- 0.1\%)-50.2\% (+/- 3.0\%) & 5.5\% (+/- 0.9\%) & 18.1\% (+/- 2.4\%)  \\
 & Census & Race & 2,901 (+/- 1,365) & 79.2\% (+/- 0.0\%)-84.7\% (+/- 0.3\%) & 0.4\% (+/- 0.1\%)-23.4\% (+/- 2.5\%) & 3.5\% (+/- 0.7\%) & 10.4\% (+/- 2.1\%) & 0.4\% (+/- 0.1\%)-38.1\% (+/- 4.0\%) & 4.5\% (+/- 1.1\%) & 16.8\% (+/- 3.9\%) \\
 & German & Sex & 77,395 (+/- 28,652) & 65.4\% (+/- 0.1\%)-76.4\% (+/- 0.3\%) & 0.0\% (+/- 0.0\%)-30.1\% (+/- 4.9\%) & 9.9\% (+/- 0.4\%) & 10.3\% (+/- 0.6\%) & 0.0\% (+/- 0.0\%)-47.5\% (+/- 2.4\%) & 10.8\% (+/- 2.6\%) & 12.1\% (+/- 3.5\%)  \\
 & Bank & Age & 3,512 (+/- 569) & 87.1\% (+/- 0.0\%)-89.4\% (+/- 0.2\%) & 0.0\% (+/- 0.0\%)-5.9\% (+/- 1.0\%) & 0.2\% (+/- 0.1\%) & 3.9\% (+/- 0.8\%) & 0.0\% (+/- 0.0\%)-10.8\% (+/- 1.8\%) & 0.2\% (+/- 0.1\%) & 7.3\% (+/- 1.6\%) \\
& Compas & Sex & 29,916 (+/- 3,149) & 92.8\% (+/- 0.0\%)-97.1\% (+/- 0.0\%) & 0.5\% (+/- 0.2\%)-5.7\% (+/- 0.3\%) & 0.8\% (+/- 0.1\%) & 4.5\% (+/- 0.7\%) & 0.0\% (+/- 0.0\%)-7.0\% (+/- 0.5\%) & 0.0\% (+/- 0.0\%)	& 4.9\% (+/- 1.4\%) \\
& Compas & Race & 29,961 (+/- 3,2) & 92.8\% (+/- 0.0\%)-97.1\% (+/- 0.0\%) & 0.8\% (+/- 0.1\%)-4.8\% (+/- 0.2\%) & 0.8\% (+/- 0.1\%) & 2.4\% (+/- 0.2\%) & 0.0\% (+/- 0.0\%)-6.0\% (+/- 0.8\%) & 0.0\% (+/- 0.0\%) & 1.6\% (+/- 0.5\%) \\
\hline
\multirow{6}{*}{DA}  
 & Census & Sex & 12,613 (+/- 3867) & 79.1\% (+/- 0.0\%)-80.2\% (+/- 0.0\%) & 0.9\% (+/- 0.0\%)-14.8\% (+/- 0.0\%) & 0.9\% (+/- 0.0\%) & 11.1\% (+/- 0.0\%) & 0.0\% (+/- 0.0\%)-24.0\% (+/- 0.0\%) & 0.0\% (+/- 0.0\%) & 13.7\% (+/- 0.0\%) \\
 & Census & Race & 7,427 (+/- 1,375) & 79.1\% (+/- 0.0\%)-80.1\% (+/- 0.0\%) & 4.8\% (+/- 0.0\%)-15.1\% (+/- 0.0\%) & 4.8\% (+/- 0.0\%) & 15.0\% (+/- 0.0\%) & 6.4\% (+/- 0.1\%)-21.1\% (+/- 0.0\%) & 6.4\% (+/- 0.0\%) & 20.9\% (+/- 0.1\%) \\
 & Credit & Sex & 62,917 (+/- 12,349) & 72.8\% (+/- 0.0\%)-77.6\% (+/- 0.0\%) & 0.2\% (+/- 0.0\%)-17.7\% (+/- 0.0\%) & 2.5\% (+/- 0.0\%) & 13.1\% (+/- 0.0\%) & 0.5\% (+/- 0.1\%)-22.8\% (+/- 0.0\%) & 3.3\% (+/- 0.0\%) & 20.0\% (+/- 0.0\%)  \\
 & Bank & Age & 2,786 (+/- 507) & 81.1\% (+/- 0.3\%)-89.2\% (+/- 0.0\%) & 0.2\% (+/- 0.0\%)-5.4\% (+/- 0.0\%) & 0.3\% (+/- 0.0\%) & 5.4\% (+/- 0.0\%) & 0.1\% (+/- 0.0\%)-10.5\% (+/- 0.0\%) & 0.4\% (+/- 0.3\%) & 10.5\% (+/- 0.0\%)  \\
 & Compas & Sex & 45,448 (+/- 95) & 96.1\% (+/- 0.0\%)-97.1\% (+/- 0.0\%) & 1.6\% (+/- 0.0\%)-3.1\% (+/- 0.0\%) & 1.6\% (+/- 0.0\%) & 3.1\% (+/- 0.0\%) & 0.0\% (+/- 0.0\%)-0.8\% (+/- 0.0\%) & 0.0\% (+/- 0.0\%)	& 0.8\% (+/- 0.0\%) \\
 & Compas & Race & 45,173 (+/- 746) & 96.1\% (+/- 0.0\%)-97.1\% (+/- 0.0\%) & 1.5\% (+/- 0.0\%)-3.1\% (+/- 0.0\%) & 1.5\% (+/- 0.0\%) & 3.1\% (+/- 0.0\%) & 0.0\% (+/- 0.0\%)-0.5\% (+/- 0.0\%) & 0.0\% (+/- 0.0\%)	& 0.5\% (+/- 0.0\%) \\
\hline

\end{tabu}
}
\label{table:bias-magnitude}
\end{table*}
}

%% file: RQ-2-Table.tex
{
\scriptsize
\begin{table*}[ht]
\caption{The performance of different search strategies to find biases in ML libraries
(discrepancies are highlighted by red).}
\centering
\resizebox{1.0\textwidth}{!}{%
\begin{tabu}{|l|l|l|lll|lll|lll|}
\hline
\multirow{2}{*}{\textbf{Algorithm}} & \multirow{2}{*}{\textbf{Dataset}} & \multirow{2}{*}{\textbf{Protected}} & \multicolumn{3}{c|}{\textbf{Num. Inputs}} & \multicolumn{3}{c|}{\textbf{|~AOD.max()~-~AOD.min()~|}} & \multicolumn{3}{c|}{\textbf{|~EOD.max()~-~EOD.min()~|}} \\
 &  &  & \textit{Random} & \textit{Black-Box} & \textit{Gray-Box} & \textit{Random} & \textit{Black-Box} & \textit{Gray-Box}
 & \textit{Random} & \textit{Black-Box} & \textit{Gray-Box}  \\
\hline


\multirow{6}{*}{LR} & Census & Sex &
11,469 (+/- 5,282) & 10,915 (+/- 5,416) & 12,763 (+/- 5,768) & 12.2\% (+/- 1.1\%) & 11.8\% (+/- 0.4\%) & 12.4\% (+/- 1.6\%) & 23.0\% (+/- 0.0\%) & 23.0\% (+/- 0.0\%) & 23.0\% (+/- 0.0\%) \\
 & Census & Race &  6,402 (+/- 2,602) & 6,592 (+/- 2,327) & 7,050 (+/- 2,551) & 13.6\% (+/- 2.4\%) & 14.2\% (+/- 2.5\%) & 15.0\% (+/- 2.6\%) & 19.4\% (+/- 3.5\%) & 20.0\% (+/- 3.5\%) & 21.5\% (+/- 3.7\%) \\
 & Credit & Sex & 34,217 (+/- 13,821) & 34,394 (+/- 13,878) &	24,248 (+/-15,175) &	12.7\% (+/- 0.3\%) & 12.6\% (+/- 0.5\%) & 12.1\% (+/- 0.8\%) & 24.9\% (+/- 1.1\%) & 25.1\% (+/- 1.1\%) & 23.9\% (+/- 1.4\%)  \\
 & Bank & Age & 2,201\% (+/- 462) &	2,267 (+/- 801) &	2,676 (+/- 982) &	8.9\% (+/- 0.0\%) & 8.8\% (+/- 0.3\%) & 8.8\% (+/- 0.2\%) & 15.1\% (+/- 0.0\%) & 14.9\% (+/- 0.4\%) & 14.9\% (+/- 0.4\%) \\
 & Compas & Sex & 70,452 (+/- 1,686) & 70,737 (+/- 2,268) & 62,020 (+/- 1,955) & 3.8\% (+/- 0.4\%) & 3.6\% (+/- 0.2\%) & 3.8\% (+/- 0.4\%) & 6.3\% (+/- 1.1\%) & 5.9\% (+/- 0.6\%) & 6.3\% (+/- 1.1\%) \\
 & Compas & Race & 70,068 (+/- 1,680) & 66,862 (+/- 9,320) & 61,755 (+/- 3,012) & 2.8\% (+/- 0.1\%) & 2.8\% (+/- 0.0\%) & 2.8\% (+/- 0.1\%) & 5.2\% (+/- 0.4\%) & 5.0\% (+/- 0.0\%) & 5.2\% (+/- 0.4\%)\\
\hline
\multirow{6}{*}{RF} & Census & Sex & 623 (+/- 247) &	603 (+/- 217) &	694 (+/- 172) &	13.3\% (+/- 1.3\%) & 13.5\% (+/- 1.5\%) & 13.4\% (+/- 1.4\%) & 27.6\% (+/- 1.2\%) & 27.5\% (+/- 1.3\%) & 27.3\% (+/- 1.4\%) \\
 & Census & Race & 575 (+/- 218) &	681 (+/- 232) &	777 (+/- 653) &	7.4\% (+/- 0.6\%) & 7.0\% (+/- 0.8\%) & 6.8\% (+/- 0.5\%) & 13.5\% (+/- 1.1\%) & 13.1\% (+/- 1.6\%) & 12.2\% (+/- 1.0\%) \\
 & Credit & Sex & 41,737 (+/- 15,717) & 39,221 (+/- 12,974) &	10,151 (+/- 3,458) &	14.9\% (+/- 0.3\%) & 15.2\% (+/- 0.6\%) & 14.8\% (+/- 0.6\%) & 24.5\% (+/- 0.7\%) & 25.0\% (+/- 0.8\%) & 23.9\% (+/- 1.2\%) \\
 & Bank & Age & 260 (+/- 133) &	314 (+/- 102) &	649 (+/- 445) &	3.1\% (+/- 0.2\%) & 3.0\% (+/- 0.5\%) & 3.0\% (+/- 0.3\%) & 5.4\% (+/- 0.4\%) & 5.7\% (+/- 0.9\%) & 5.4\% (+/- 0.5\%)  \\
  & Compas & Sex & 28,930 (+/- 803) & 29,780 (+/- 958) & 13,216 (+/- 1,050) & 3.9\% (+/- 0.3\%) & 3.7\% (+/- 0.4\%) & 3.4\% (+/- 0.2\%) & 7.6\% (+/- 0.6\%) & 7.1\% (+/- 0.7\%) & 6.6\% (+/- 0.4\%) \\
 & Compas & Race & 27,580 (+/- 4,052) & 27,873 (+/- 3,948) & 12,426 (+/- 1,122) & 3.2\% (+/- 0.2\%) & 3.3\% (+/- 0.2\%) & 2.9\% (+/- 0.3\%) & 6.6\% (+/- 0.5\%) & 6.7\% (+/- 0.5\%) & 5.9\% (+/- 0.7\%) \\
\hline
\multirow{6}{*}{SVM} & Census & Sex &
95,543 (+/- 57) & 96,214.0 (+/- 5,610) & 97,115 (+/- 1,385) & 32.6\% (+/- 0.2\%) & 32.6\% (+/- 0.2\%) & 32.6\% (+/- 0.2\%) & 30.5\% (+/- 1.4\%) & 28.6\% (+/- 2.5\%) & 29.9\% (+/- 1.7\%) \\
 & Census & Race & 40,311 (+/- 502) & 41,289 (+/- 1,262) & 39,572 (+/- 869) & 31.3\% (+/- 1.6\%) & 29.5\% (+/- 1.6\%) & 30.7\% (+/- 1.2\%) & 39.3\% (+/- 2.3\%) & 36.2\% (+/- 2.6\%) & 38.0\% (+/- 2.2\%) \\
 & Credit & Sex & 5,710 (+/- 925) & 4,388 (+/- 919)	& 6,149 (+/- 980) & 18.2\% (+/- 1.5\%) & 17.5\% (+/- 0.0\%) & 17.5\% (+/- 0.0\%) & 25.0\% (+/- 0.9\%) & 24.8\% (+/- 1.6\%) & 24.7\% (+/- 1.1\%) \\
 & Bank & Age & 1,467 (+/- 108) & 1,359 (+/- 444) & 1,083 (+/- 925) & 5.6\% (+/- 0.4\%) & 4.8\% (+/- 0.6\%) & 4.7\% (+/- 0.8\%) & 9.8\% (+/- 0.5\%) & 8.5\% (+/- 1.0\%) & 8.3\% (+/- 1.3\%) \\
  & Compas & Sex & 40,070 (+/- 1,265) & 40,355 (+/- 849) & 40,401 (+/- 375) & 2.3\% (+/- 0.2\%) & 2.1\% (+/- 0.1\%) & 2.3\% (+/- 0.3\%) & 3.9\% (+/- 0.2\%) & 3.7\% (+/- 0.2\%) & 4.0\% (+/- 0.4\%) \\
 & Compas & Race & 3,911 (+/- 1,132) & 5,430 (+/- 841) & 4,426.0 (+/- 1,202.0) & 1.7\% (+/- 0.1\%) & 1.6\% (+/- 0.0\%) & 1.6\% (+/- 0.0\%) & 3.0\% (+/- 0.1\%) & 2.9\% (+/- 0.2\%) & 2.9\% (+/- 0.2\%) \\
\hline
\multirow{6}{*}{DT} & Census & Sex &
\textcolor{red}{158 (+/- 92)} &	\textcolor{red}{7,351 (+/- 1,243)} &	\textcolor{red}{5,804 (+/- 2,031)} & \textcolor{red}{26.8\% (+/- 2.3\%)} & \textcolor{red}{32.8\% (+/- 0.7\%)} & \textcolor{red}{35.1\% (+/- 3.4\%)} & \textcolor{red}{40.6\% (+/- 5.8\%)} & \textcolor{red}{52.7\% (+/- 2.0\%)} & \textcolor{red}{55.4\% (+/- 4.8\%)}  \\
  & Census & Race & \textcolor{red}{125 (+/- 9)} & \textcolor{red}{6,645 (+/- 1,949)} & \textcolor{red}{5,094 (+/- 2,250)} & \textcolor{red}{18.0\% (+/- 2.0\%)} & \textcolor{red}{29.3\% (+/- 1.2\%)} & \textcolor{red}{25.7\% (+/- 4.5\%)} & \textcolor{red}{30.1\% (+/- 3.5\%)} & \textcolor{red}{47.1\% (+/- 2.0\%)} & \textcolor{red}{41.7\% (+/- 7.5\%)} \\
 & Credit & Sex & 86,762 (+/- 15,750) & 86,588 (+/- 15,545.0) & 72,522 (+/- 27,373) & 34.4\% (+/- 5.1\%) & 30.5\% (+/- 3.7\%) & 30.8\% (+/- 3.1\%) & 51.6\% (+/- 2.2\%) & 48.2\% (+/- 6.7\%) & 49.3\% (+/- 4.3\%) \\
 & Bank & Age & \textcolor{red}{3,322 (+/- 782)} & \textcolor{red}{3,447 (+/- 1,191)} & \textcolor{red}{3,689 (+/- 1,119)} & \textcolor{red}{3.8\% (+/- 1.0\%)} & \textcolor{red}{7.3\% (+/- 1.5\%)} & \textcolor{red}{6.9\% (+/- 1.3\%)} & \textcolor{red}{6.9\% (+/- 2.0\%)} & \textcolor{red}{13.7\% (+/- 2.6\%)} & \textcolor{red}{12.6\% (+/- 2.6\%)} \\
   & Compas & Sex & \textcolor{red}{18,442 (+/- 79)} & \textcolor{red}{36,142 (+/- 330)} & \textcolor{red}{34,607 (+/- 1,102)} & \textcolor{red}{4.1\% (+/- 0.6\%)} & \textcolor{red}{6.0\% (+/- 0.3\%)} & \textcolor{red}{5.5\% (+/- 0.7\%)} & \textcolor{red}{5.5\% (+/- 0.6\%)} & \textcolor{red}{8.1\% (+/- 0.3\%)} & \textcolor{red}{7.5\% (+/- 0.8\%)} \\
 & Compas & Race & \textcolor{red}{18,512 (+/- 117)} & \textcolor{red}{36,073 (+/- 517)} & \textcolor{red}{35,502 (+/- 858)} & \textcolor{red}{2.8\% (+/- 0.5\%)} & \textcolor{red}{4.7\% (+/- 0.2\%)} & \textcolor{red}{4.6\% (+/- 0.3\%)} & \textcolor{red}{3.9\% (+/- 0.2\%)} & \textcolor{red}{7.9\% (+/- 0.9\%)} & \textcolor{red}{6.8\% (+/- 1.3\%)} \\
\hline
\multirow{6}{*}{DA} & Census & Sex & 17,553 (+/- 6,794) & 15,054 (+/- 6,993) & 12,784 (+/- 5,609) & 13.9\% (+/- 0.0\%) & 13.9\% (+/- 0.0\%) & 13.9\% (+/- 0.0\%) & 24.0\% (+/- 0.0\%) & 24.0\% (+/- 0.0\%) & 24.0\% (+/- 0.0\%) \\
 & Census & Race & 8,399 (+/- 2964) &	7,816 (+/- 2,694) & 7,051 (+/- 2408) & 10.4\% (+/- 0.0\%) & 10.4\% (+/- 0.0\%) & 10.3\% (+/- 0.1\%) & 14.7\% (+/- 0.0\%) & 14.7\% (+/- 0.0\%) & 14.7\% (+/- 0.1\%) \\
 & Credit & Sex & 67,283 (+/- 21,948) & 7,232 (+/- 22,043) & 54,237 (+/- 27,776) & 5.2\% (+/- 0.0\%) & 5.2\% (+/- 0.0\%) & 5.3\% (+/- 0.1\%) & 10.4\% (+/- 0.0\%) & 10.4\% (+/- 0.0\%) & 10.4\% (+/- 0.0\%) \\
 & Bank & Age & 2,812 (+/- 663) &	2,809 (+/- 978) & 2,878 (+/- 929) & 17.5\% (+/- 0.0\%) & 17.5\% (+/- 0.0\%) & 17.5\% (+/- 0.0\%) & 22.4\% (+/- 0.0\%) & 22.4\% (+/- 0.0\%) & 22.4\% (+/- 0.0\%)  \\
 & Compas & Sex & 45,489 (+/- 94) & 45,449 (+/- 175) & 45,406 (+/- 257) & 1.5\% (+/- 0.0\%) & 1.5\% (+/- 0.0\%) & 1.5\% (+/- 0.0\%) & 0.8\% (+/- 0.0\%) & 0.8\% (+/- 0.0\%) & 0.8\% (+/- 0.0\%) \\
 & Compas & Race & 44,436 (+/- 2432) & 45,603 (+/- 300) & 45,480 (+/- 339) & 1.6\% (+/- 0.0\%) & 1.6\% (+/- 0.0\%) & 1.6\% (+/- 0.0\%) & 0.5\% (+/- 0.0\%) & 0.5\% (+/- 0.0\%) & 0.5\% (+/- 0.0\%) \\
\hline
\end{tabu}
}
\label{table:search-bias}
\end{table*}
}

%% file: RQ-2-Figures.tex
\begin{figure*}[!htb]
    \centering
    \begin{minipage}{0.245\textwidth}
        \centering
   		\includegraphics[width=1\textwidth]{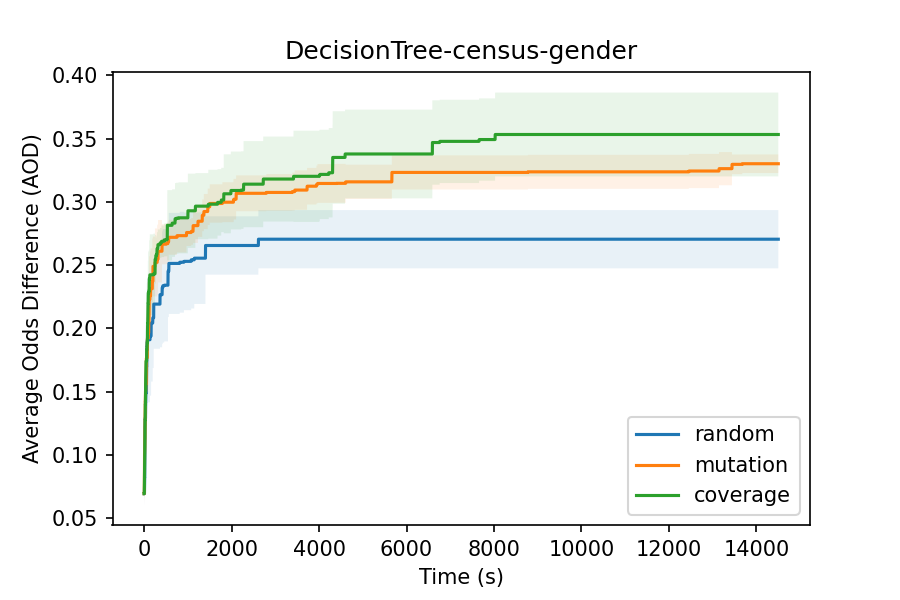}
    \end{minipage}%
    \begin{minipage}{0.245\textwidth}
        \centering
        \includegraphics[width=1\textwidth]{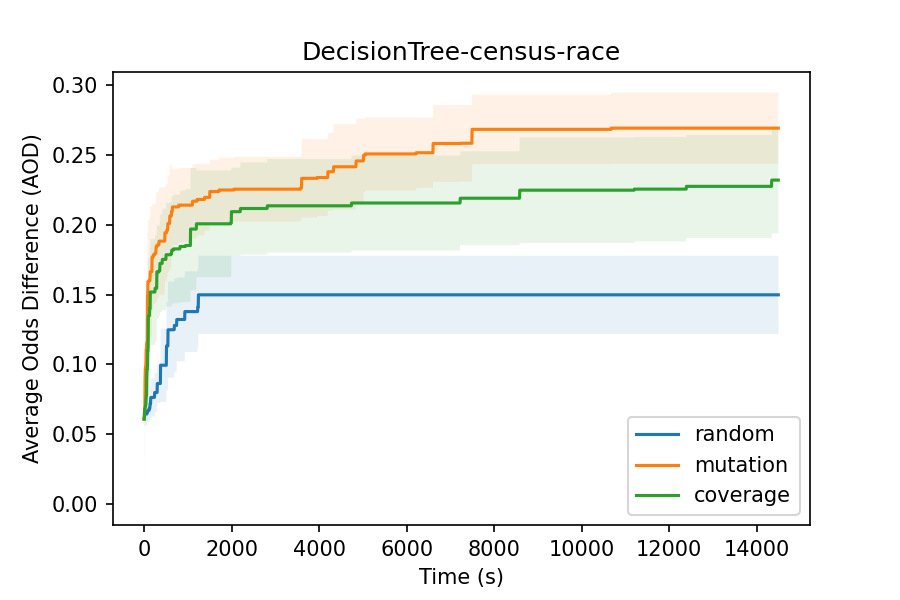}
    \end{minipage}%
    \begin{minipage}{0.245\textwidth}
    	\centering
        \includegraphics[width=1\textwidth]{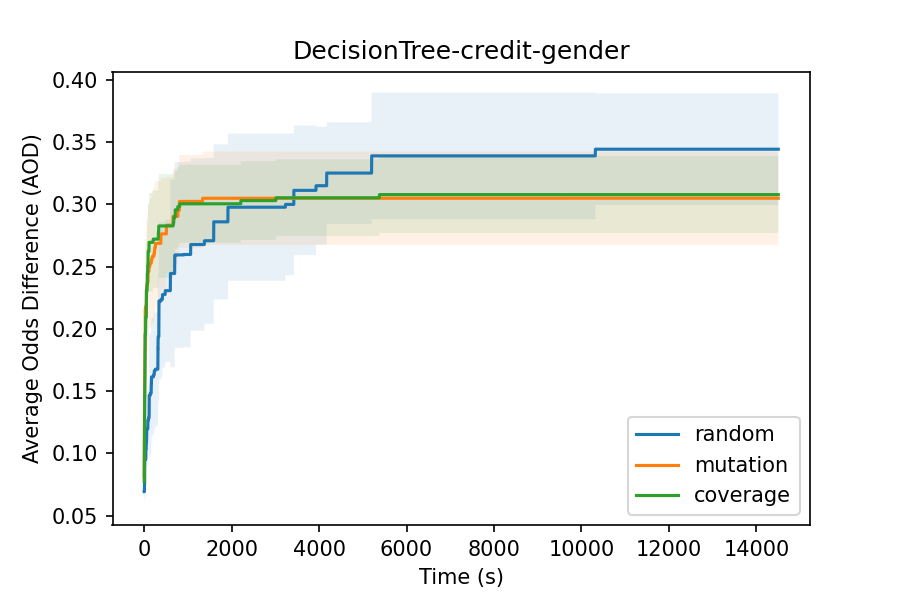}
    \end{minipage}
    \begin{minipage}{0.245\textwidth}
    	\centering
        \includegraphics[width=1\textwidth]{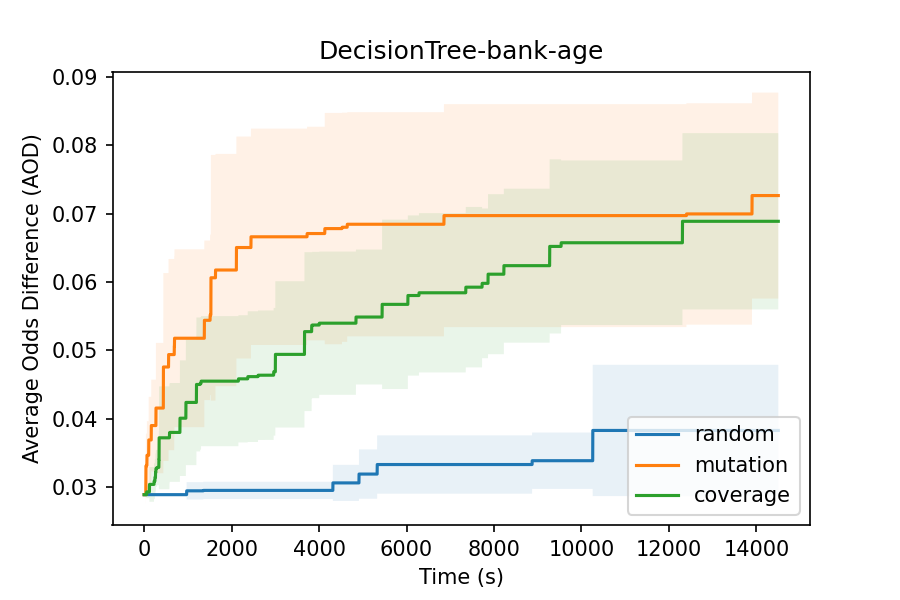}
    \end{minipage}
    \caption{
    The temporal progress of search strategies for the decision tree algorithm
    over $4$ training tasks. X-axis is the timestamp of search from 0s to 14000s (4 hrs)
    and Y-axis is the group fairness metric ($AOD$). $Coverage$ refers to gray-box
    method.
    }
    \label{fig:search-plots}
\end{figure*}

%% file: RQ-3-Figures.tex
\begin{figure*}[!htb]
  \centering
  \begin{subfigure}{0.24\textwidth}
    \includegraphics[width=1\textwidth]{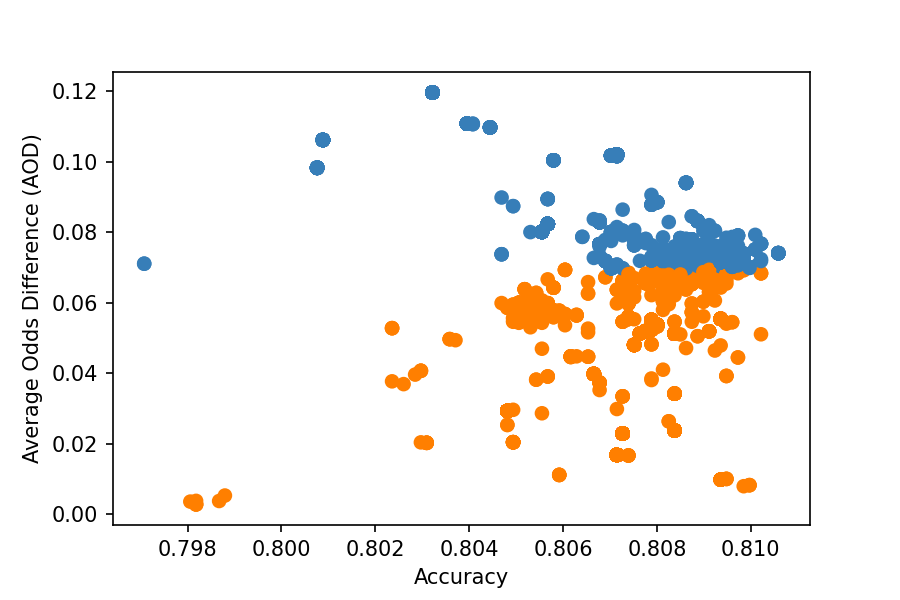}
    \caption{Logistic Regression Cluters.}
  \end{subfigure}
  \hfill
  \begin{subfigure}{0.24\textwidth}
    \includegraphics[width=0.68\textwidth]{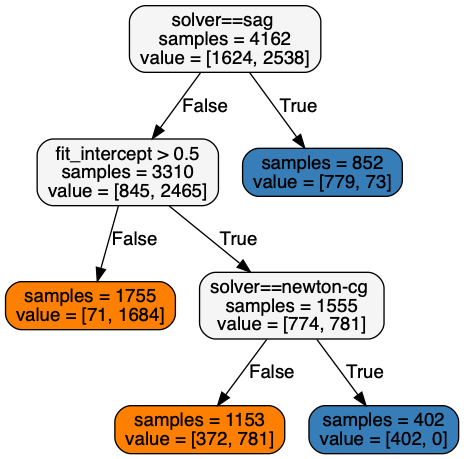}
    \caption{Logistic Regression Explained.}
  \end{subfigure}
  \hfill
  \begin{subfigure}{0.24\textwidth}
    \includegraphics[width=1\textwidth]{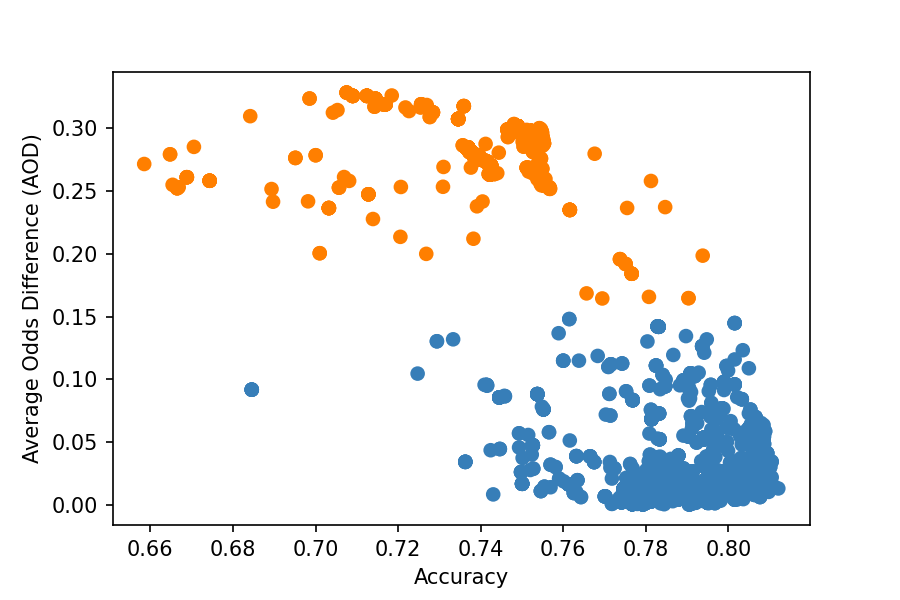}
    \caption{SVM Clusters.}
  \end{subfigure}
  \hfill
  \begin{subfigure}{0.25\textwidth}
    \includegraphics[width=0.9\textwidth]{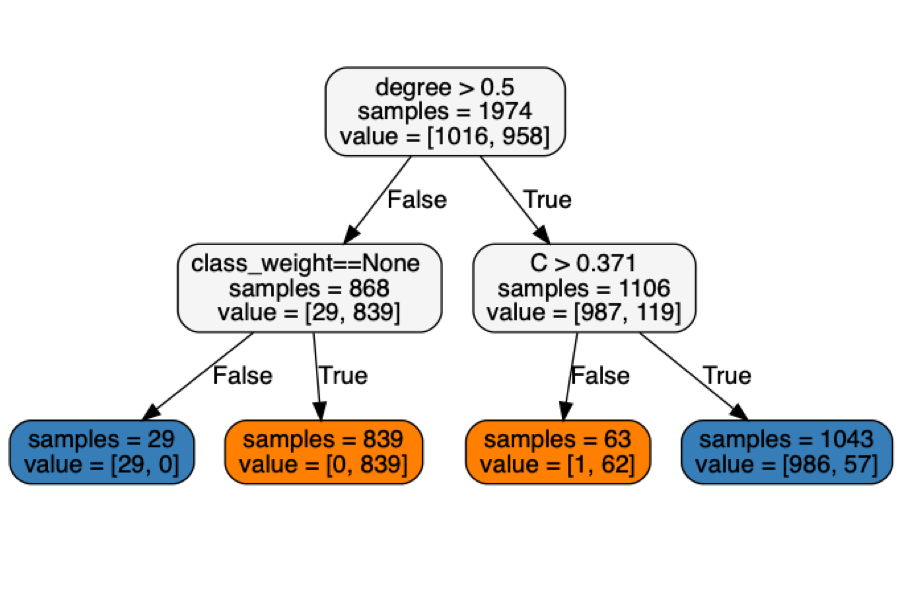}
    \caption{SVM Explained.}
  \end{subfigure}
  \hfill
  \begin{subfigure}{0.24\textwidth}
    \includegraphics[width=1\textwidth]{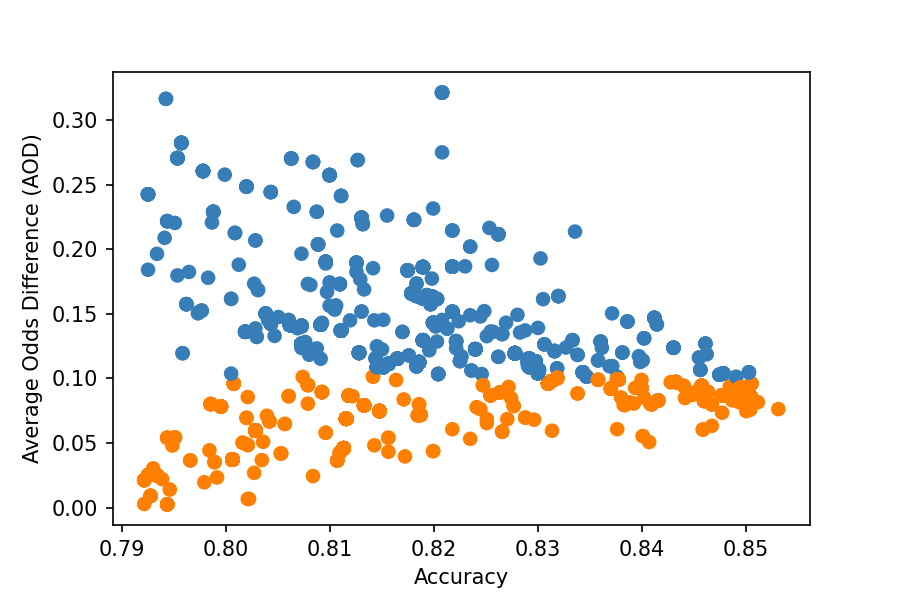}
    \caption{Decision Tree Clusters.}
  \end{subfigure}
  \hfill
  \begin{subfigure}{0.25\textwidth}
    \includegraphics[width=0.85\textwidth]{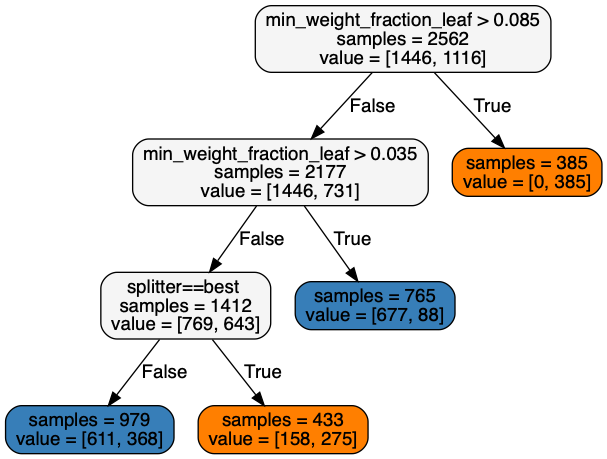}
    \caption{Decision Tree Explained.}
  \end{subfigure}
  \hfill
  \begin{subfigure}{0.24\textwidth}
    \includegraphics[width=1\textwidth]{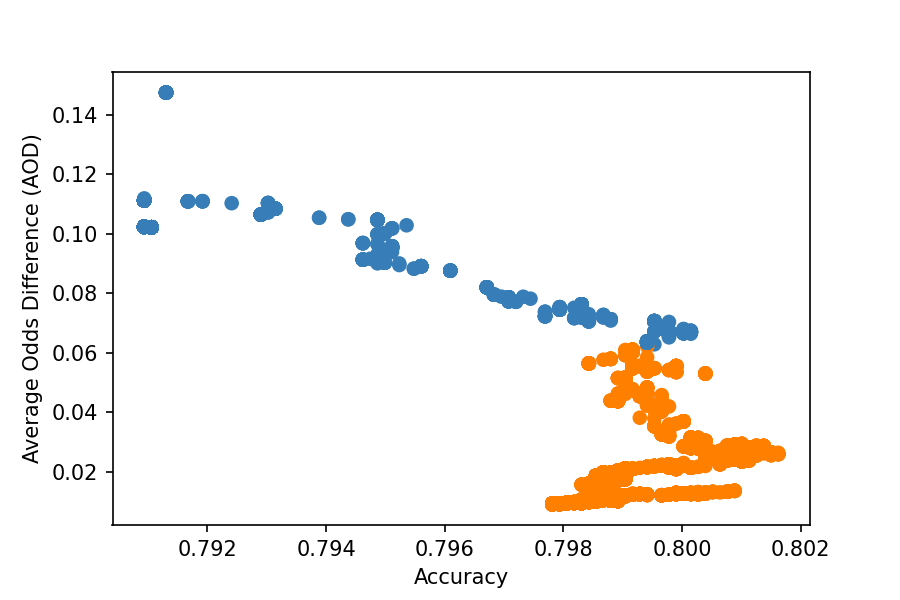}
    \caption{Discriminant Clusters.}
  \end{subfigure}
  \hfill
  \begin{subfigure}{0.24\textwidth}
    \includegraphics[width=0.58\textwidth]{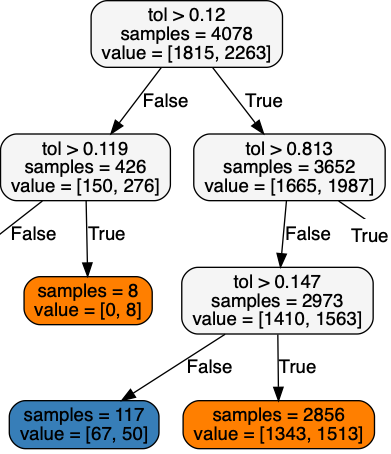}
    \caption{Discriminant Explained.}
  \end{subfigure}
\caption{The test inputs over census dataset with sex as the protected attribute are (1) clustered
into two groups in the domain of fairness and accuracy (2) explained to understand which
parameters distinguish low and high fairness outcomes.}
\label{fig:clustering-explanation}
\end{figure*}

%% file: RQ-4-Table.tex
{
\scriptsize
\begin{table*}[ht]
\caption{\toolname as a bias mitigation technique compared to Exp. Gradient~\cite{agarwal2018reductions} within 6 mins.}
\centering
\resizebox{1.0\textwidth}{!}{%
\begin{tabu}{|l|l|l|ll|ll|ll|ll|}
\hline
\multirow{2}{*}{\textbf{Algorithm}} & \multirow{2}{*}{\textbf{Dataset}} & \multirow{2}{*}{\textbf{Protected}} & \multicolumn{2}{c|}{\textbf{Default Configuration}} & \multicolumn{2}{c|}{\textbf{Exp. Gradient~\cite{agarwal2018reductions}}} & \multicolumn{2}{c|}{\textbf{\toolname}}
 & \multicolumn{2}{c|}{\textbf{\toolname} + \textbf{Exp. Gradient~\cite{agarwal2018reductions}}} \\
 &  &  & \textit{Accuracy} & \textit{EOD} & \textit{Accuracy} & \textit{EOD} & \textit{Accuracy} & \textit{EOD} & \textit{Accuracy} & \textit{EOD} \\
\hline


\multirow{6}{*}{LR}
 & Census & Sex  & 80.5\% (+/- 0.0\%) & 9.7\% (+/- 0.1\%) & 80.5\% (+/- 0.1\%) & \textcolor{red}{0.8\% (+/- 0.3\%)} & 80.2\% (+/- 0.3\%) & \textcolor{red}{0.1\% (+/- 0.0\%)} & 80.0\% (+/- 0.5\%) &  0.2\% (+/- 0.1\%) \\
 & Census & Race & 80.5\% (+/- 0.0\%) & 9.9\% (+/- 0.0\%) & 79.6\% (+/- 0.1\%) & 3.5\% (+/- 2.2\%) & 80.2\% (+/- 0.3\%) & 1.1\% (+/- 1.0\%) & 80.0\% (+/- 0.5\%) & 1.3\% (+/- 0.9\%)  \\
 & Credit & Sex  & 74.4\% (+/- 0.0\%) & 17.1\% (+/- 0.0) & 74.5\% (+/- 0.6\%) &  \textcolor{red}{25\% (+/- 1.9\%)} & 75.2\% (+/- 0.8\%) & \textcolor{red}{0.6\% (+/- 0.5\%)} & 74.3\% (+/- 0.5\%) & 1.6\% (+/- 1.2\%) \\
 & Bank   & Age  & 89.0\% (+/- 0.0\%) & 8.0\% (+/- 0.0\%) & 86.7\% (+/- 0.1\%) & \textcolor{red}{2.4\% (+/- 1.6\%)} & 88.6\% (+/- 0.2\%) & \textcolor{red}{0.0\% (+/- 0.0\%)} & 88.2\% (+/- 0.4\%) &  0.8\% (+/- 0.8\%) \\
 & Compas & Sex  & 97.0\% (+/- 0.0\%) & 1.6\% (+/- 0.0\%) & 96.9\% (+/- 0.1\%) & 0.0\% (+/- 0.0\%) & 97.1\% (+/- 0.0\%) & 0.0\% (+/- 0.0\%) & 97.1\% (+/- 0.0\%) & 0.0\% (+/- 0.0\%) \\
 & Compas & Race & 97.0\% (+/- 0.0\%) & 0.3\% (+/- 0.0\%) & 96.9\% (+/- 0.1\%) & 0.0\% (+/- 0.0\%) & 97.1\% (+/- 0.0\%) & 0.0\% (+/- 0.0\%) & 97.1\% (+/- 0.0\%) & 0.0\% (+/- 0.0\%) \\
\hline
\multirow{6}{*}{RF}
& Census & Sex  & 84.0\% (+/- 0.0\%) & 5.4\% (+/- 0.0\%) & 79.0\% (+/- 0.0\%) & 5.4\% (+/- 0.0\%) & 84.0\% (+/- 0.4\%) & 5.0\% (+/- 1.0\%) & 79.4\% (+/- 0.0\%) & 0.1\% (+/- 0.0\%)  \\
& Census & Race & 84.0\% (+/- 0.0\%) & 8.6\% (+/- 0.0\%) & 79.8\% (+/- 0.0\%) & 0.3\% (+/- 0.0\%) & 84.7\% (+/- 0.6\%) & 4.5\% (+/- 0.5\%) & 79.8\% (+/- 0.0\%) & 0.3\% (+/- 0.0\%)  \\
& Credit & Sex  & 74.0\% (+/- 0.0\%) & 11.6\% (+/- 0.0\%) & 70.4\% (+/- 0.0\%) & \textcolor{red}{8.3\% (+/- 0.0\%)} & 78.1\% (+/- 0.4\%) & \textcolor{red}{0.1\% (+/- 0.0\%)}  & 70.8\% (+/- 0.0\%) & 0.5\% (+/- 0.0\%)  \\
& Bank   & Age  & 89.9\% (+/- 0.0\%) & 1.2\% (+/- 0.0\%) & 79.0\% (+/- 0.3\%) & \textcolor{red}{3.6\% (+/- 1.6\%)} & 89.9\% (+/- 0.2\%) & \textcolor{red}{0.1\% (+/- 0.1\%)} & 83.1\% (+/- 0.8\%) & 0.7\% (+/- 0.7\%)  \\
& Compas & Sex  & 96.5\% (+/- 0.0\%) & 2.3\% (+/- 0.0\%) & 93.6\% (+/- 0.0\%) & \textcolor{red}{1.5\% (+/- 0.0\%)} & 97.1\% (+/- 0.0\%) & \textcolor{red}{0.0\% (+/- 0.0\%)} & 97.1\% (+/- 0.0\%) & 0.0\% (+/- 0.0\%)  \\
& Compas & Race &  96.5\% (+/- 0.0\%) & 2.1\% (+/- 0.0\%) & 93.7\% (+/- 0.0\%) & \textcolor{red}{1.1\% (+/- 0.0\%)} & 97.1\% (+/- 0.0\%) & \textcolor{red}{0.0\% (+/- 0.0\%)} & 97.1\% (+/- 0.0\%) & 0.0\% (+/- 0.0\%)  \\
\hline
\multirow{6}{*}{SVM}
& Census & Sex  & 66.5\% (+/- 0.0\%) & 18.5\% (+/- 0.0\%) & 79.9\% (+/- 0.0\%) & \textcolor{red}{0.7\% (+/- 0.2\%)} & 79.1\% (+/- 0.7\%) & \textcolor{red}{0.0\% (+/- 0.0\%)} & 80.1\% (+/- 0.4\%) & 0.2\% (+/- 0.1\%)  \\
& Census & Race & 72.5\% (+/- 0.0\%) & 4.5\% (+/- 0.0\%) & 72.5\% (+/- 0.4\%) & \textcolor{red}{3.8\% (+/- 1.8\%)} & 78.8\% (+/- 2.2\%) & \textcolor{red}{0.0\% (+/- 0.0\%)} & 76.3\% (+/- 1.7\%) & 0.3\% (+/- 0.3\%)   \\
& Credit & Sex  & 62.8\% (+/- 0.0\%) & 22.3\% (+/- 0.3\%) & 64.6\% (+/- 1.7\%) & \textcolor{red}{10.4\% (+/- 0.0\%)} & 70.4\% (+/- 0.0\%) & \textcolor{red}{0.0\% (+/- 0.0\%)} & 70.4\% (+/- 0.0\%) & 0.0\% (+/- 0.0\%)  \\
& Bank   & Age  &  89.9\% (+/- 0.0\%) & 1.2\% (+/- 0.0\%) & 79.0\% (+/- 0.3\%) & \textcolor{red}{3.6\% (+/- 1.6\%)} & 89.2\% (+/- 0.3\%) & \textcolor{red}{0.0\% (+/- 0.0\%)} & 88.3\% (+/- 0.1\%) & 0.0\% (+/- 0.0\%)  \\
& Compas & Sex  & 96.5\% (+/- 0.0\%) & 0.0\% (+/- 0.0\%) & 93.6\% (+/- 0.0\%) & 0.0\% (+/- 0.0\%) & 97.1\% (+/- 0.0\%) & 0.0\% (+/- 0.0\%) & 97.1\% (+/- 0.0\%) & 0.0\% (+/- 0.0\%)  \\
& Compas & Race & 96.5\% (+/- 0.0\%) & 0.0\% (+/- 0.0\%) & 93.7\% (+/- 0.0\%) & 0.0\% (+/- 0.0\%) & 97.1\% (+/- 0.0\%) & 0.0\% (+/- 0.0\%) & 97.1\% (+/- 0.0\%) & 0.0\% (+/- 0.0\%)  \\
\hline
\multirow{6}{*}{DT}
& Census & Sex  & 80.2\% (+/- 0.0\%) & 3.3\% (+/- 0.0\%) & 82.6\% (+/- 0.1\%) & 2.1\% (+/- 0.6\%) & 79.9\% (+/- 0.8\%) & 0.2\% (+/- 0.2\%) & 82.7\% (+/- 0.5\%) & 0.9\% (+/- 1.0\%)  \\
& Census & Race & 80.2\% (+/- 0.0\%) & 7.2\% (+/- 0.0\%) & 83.0\% (+/- 0.1\%) & 7.5\% (+/- 1.6\%) & 80.7\% (+/- 0.9\%) & 0.2\% (+/- 0.1\%) & 83.2\% (+/- 0.5\%) & 1.7\% (+/- 1.0\%)  \\
& Credit & Sex  &  66.0\% (+/- 0.0\%) & 13.4\% (+/- 0.0\%) &  70.0\% (+/- 0.3\%) &  \textcolor{red}{14.9\% (+/- 1.8\%)} & 70.4\% (+/- 0.0\%) & \textcolor{red}{0.0\% (+/- 0.0\%)} & 70.4\% (+/- 0.0\%) & 0.0\% (+/- 0.0\%)  \\
& Bank   & Age  & 87.1\% (+/- 0.0\%) & 4.8\% (+/- 0.0\%) & 88.0\% (+/- 0.1\%) & \textcolor{red}{3.7\% (+/- 1.7\%)} & 88.4\% (+/- 0.3\%) & \textcolor{red}{0.0\% (+/- 0.0\%)} & 88.3\% (+/- 0.0\%) & 0.0\% (+/- 0.0\%)  \\
& Compas & Sex  & 93.8\% (+/- 0.0\%) & 5.2\% (+/- 0.0\%) & 95.9\% (+/- 0.0\%) & 2.3\% (+/- 0.0\%) & 97.1\% (+/- 0.0\%) & 0.0\% (+/- 0.0\%) & 97.1\% (+/- 0.0\%) & 0.0\% (+/- 0.0\%)  \\
& Compas & Race & 93.8\% (+/- 0.0\%) & 3.8\% (+/- 0.0\%) & 96.0\% (+/- 0.0\%) & 1.3\% (+/- 0.0\%) & 97.1\% (+/- 0.0\%) & 0.0\% (+/- 0.0\%) & 97.0\% (+/- 0.1\%) & 0.0\% (+/- 0.0\%)  \\
\hline
\end{tabu}
}
\label{table:comparison}
\end{table*}
}

%% file: RQ-4-Table-2.tex
{\footnotesize
\begin{table*}[t]
\caption{\toolname in comparison to \textsc{Fairway} (\textsc{FLASH})~\cite{chakraborty2019software,chakraborty2020fairway}.}
\centering
\resizebox{0.75\textwidth}{!}{%
\begin{tabu}{|l|l|l|lll|lll|}
\hline
\multirow{2}{*}{\textbf{Alg.}} & \multirow{2}{*}{\textbf{Scenario}} &
\textbf{Time} & \multicolumn{3}{c|}{\textbf{\textsc{FLASH}}} & \multicolumn{3}{c|}{\textbf{\toolname}} \\
 &  & ~~~~~~~~~~~~~~(s) & \textit{Accuracy} & \textit{AOD} & \textit{EOD} & \textit{Accuracy} & \textit{AOD} & \textit{EOD} \\
\hline


\multirow{6}{*}{LR}
 & Census,~Sex  &  40.3 & 80.5\% (+/- 0.1\%) &	2.0\% (+/- 0.1\%)	& \textcolor{red}{0.2\% (+/- 0.3\%)} & 80.9\% (+/- 0.2\%)	& 2.7\% (+/- 1.0\%) & \textcolor{red}{4.3\% (+/- 1.3\%)} \\
 & Census,Race & 65.2 & 80.3\% (+/- 0.0\%) & \textcolor{red}{\textbf{6.2\% (+/- 0.4\%)}} &	\textcolor{red}{8.0\% (+/- 0.6\%)} & 80.9\% (+/- 0.0\%) &	\textcolor{red}{\textbf{4.2\% (+/- 0.8\%)}} & \textcolor{red}{5.8\% (+/- 1.4\%)} \\
 & Credit,~Sex  & 2.8 &	70.4\% (+/- 0.0\%) & 0.0\% (+/- 0.0\%) & 0.0\% (+/- 0.0\%) & 76.1\% (+/- 0.0\%) & 3.5\% (+/- 0.0\%) & 5.5\% (+/- 0.0\%) \\
 & Bank,~Age & 129.0 &  90.5\% (+/- 0.0\%) & 0.8\% (+/- 0.1\%) & 1.1\% (+/- 0.2\%) &
 89.6\% (+/- 0.0\%) &	0.4\% (+/- 0.2\%) & 0.4\% (+/- 0.0\%) \\
 & Compas,~Sex  & 18.3 &	97.1\% (+/- 0.0\%) & 1.6\% (+/- 0.0\%)	& 0.0\% (+/- 0.0\%) & 97.1\% (+/- 0.0\%)	& 1.6\% (+/- 0.0\%) & 0.0\% (+/- 0.0\%) \\
 & Compas,~Race & 9.5 &	97.1\% (+/- 0.0\%) & 1.5\% (+/- 0.0\%) &	0.0\% (+/- 0.0\%) & 97.1\% (+/- 0.0\%) & 1.5\% (+/- 0.0\%) & 0.0\% (+/- 0.0\%) \\
\hline
\multirow{6}{*}{DT}
& Census,~Sex  & 25.8 & 82.5\% (+/- 0.8\%) & 7.9\% (+/- 2.2\%) &	\textcolor{red}{9.9\% (+/- 3.7\%)} & 83.6\% (+/- 0.8\%) & 4.8\% (+/- 1.1\%) & \textcolor{red}{2.2\% (+/- 0.5\%)} \\
& Census,~Race & 60.5 & 83.9\% (+/- 0.7\%) & 3.2\% (+/- 1.0\%) &	4.3\% (+/- 1.8\%) & 84.1\% (+/- 0.8\%) & 3.1\% (+/- 1.1\%) & 4.4\% (+/- 1.6\%) \\
& Credit,~Sex  &  2.5 &	71.5\% (+/- 1.3\%) & \textcolor{red}{\textbf{5.4\% (+/- 2.2\%)}} &	\textcolor{red}{7.7\% (+/- 4.2\%)} &  71.1\% (+/- 0.0\%) & \textcolor{red}{\textbf{0.0\% (+/- 0.0\%)}} & \textcolor{red}{0.0\% (+/- 0.0\%)} \\
 & Bank,~Age & 124.8 & 90.9\% (+/- 0.1\%) & \textcolor{red}{\textbf{0.7\% (+/- 0.4\%)}} & \textcolor{red}{0.8\% (+/- 0.8\%)} & 88.9\% (+/- 0.4\%) &	\textcolor{red}{\textbf{0.0\% (+/- 0.0\%)}} & \textcolor{red}{5.6\% (+/- 0.0\%)} \\
& Compas,~Sex  & 5.9 &	97.1\% (+/- 0.0\%) & 1.6\% (+/- 0.0\%) &	0.0\% (+/- 0.0\%) & 97.1\% (+/- 0.0\%) & 1.5\% (+/- 0.2\%) & 0.0\% (+/- 0.0\%)   \\
& Compas,~Race & 11.4 & 97.1\% (+/- 0.0\%) & 1.5\% (+/- 0.0\%) & 0.0\% (+/- 0.0\%) & 97.1\% (+/- 0.0\%) & 1.5\% (+/- 0.0\%) & 0.0\% (+/- 0.0\%)  \\
\hline
\end{tabu}
}
\label{table:comparison-FLASH}
\end{table*}
}

%% file: discussion.tex
\vspace{0.5em}
\noindent \textit{Limitation}. The input dataset is arguably
the main source of discriminations in data-driven software.
In this work, we vary the configuration of learning algorithms
and fix the input dataset since our approach is to systematically
study the influence of hyperparameters in fairness.
While we found that configurations can reduce biases in various algorithms,
our approach alone cannot eliminate fairness bugs. Our approach also requires
a diverse set of inputs generated automatically using the search algorithms.
As a dynamic analysis, 
our approach solely relies on
heuristics to generate a diverse set of configurations and is not guaranteed
to always find interesting hyperparameters in a given time limit. In addition,
we only use two group fairness metrics ($AOD$ and $EOD$) and the overall prediction
accuracy. One limitation is that these metrics do not consider the
distribution of different groups. In general, coming up with a suitable fairness
definition is an open challenge.

\vspace{0.5em}
\noindent \textit{Threat to Validity}. To address the internal validity and ensure
our finding does not lead to invalid conclusion, we follow established
guideline~\cite{https://doi.org/10.1002/stvr.1486} where we repeat
every experiment $10$ times, report the average with $95\%$ confidence intervals (CI),
and consider not only the final result but also the temporal progresses.
We note that $95\%$ non-overlapping CI is a conservative statistical method
to compare results. Instead,
non-parametric methods and effect sizes can be used to alleviate the
conservativeness of our comparisons. 
In our experiments, we did not find significant improvements using coverage
metrics. However, this might be a result of our specific implementations and/or
the feedback criteria.
To ensure that our results are generalizable and address external validity,
we perform our experiments on five learning algorithms from scikit-learn library
over six fairness-sensitive applications that have been widely used in the fairness
literature. However, it is an open problem whether the library, algorithms, and
applications are sufficiently representative to
cover challenging fairness scenarios.

\vspace{0.5em}
\noindent \textit{Usage Vision}. \toolname complements the workflow of standard
testing procedures against functionality and performance by enabling ML library
maintainers to detect and debug fairness bugs. \toolname combines search-based
software testing with statistical debugging to identify and explain
hyperparameters that lead to high and low bias classifiers within
acceptable accuracy. Like standard ML code testing, \toolname
requires a set of reference fairness-sensitive datasets.
If the explanatory models are consistent across these datasets, \toolname
synthesizes this information to pinpoint dataset-agnostic fairness bugs.
If such bugs are discovered, ML library maintainers can either exclude
those options or warn users to avoid setting them for
fairness-sensitive applications.

%% file: conclusion.tex
Software developers increasingly employ machine learning libraries to design
data-driven social-critical applications that demand a delicate balance between
accuracy and fairness. The ``programming'' task in designing such systems
involves carefully selecting hyperparameters for these libraries, often resolved
by rules-of-thumb. We propose a search-based software engineering approach to
exploring the space of hyperparameters to approximate the twined Pareto curves
expressing both high and low fairness against accuracy.
Hyperparameter configurations with high fairness help software engineers
mitigate bias, while configurations with low fairness help ML developers
understand and document potentially unfair combinations of hyperparameters.
There are multiple exciting future directions. For example, extending our
methodology to support deep learning frameworks is an interesting future work.

%% file: ack.tex
\begin{acks}
The authors thank the anonymous reviewers for their time and
invaluable feedback to improve this paper. 
This work utilized resources from the CU Boulder Research
Computing Group, which is supported by NSF, CU Boulder, and CSU.
Tizpaz-Niari was partially supported by NSF under grant DGE-2043250 and
UTEP College of Engineering under startup package.
\end{acks}

%% file: main.bbl

\begin{thebibliography}{41}


\ifx \showCODEN    \undefined \def \showCODEN     #1{\unskip}     \fi
\ifx \showDOI      \undefined \def \showDOI       #1{#1}\fi
\ifx \showISBNx    \undefined \def \showISBNx     #1{\unskip}     \fi
\ifx \showISBNxiii \undefined \def \showISBNxiii  #1{\unskip}     \fi
\ifx \showISSN     \undefined \def \showISSN      #1{\unskip}     \fi
\ifx \showLCCN     \undefined \def \showLCCN      #1{\unskip}     \fi
\ifx \shownote     \undefined \def \shownote      #1{#1}          \fi
\ifx \showarticletitle \undefined \def \showarticletitle #1{#1}   \fi
\ifx \showURL      \undefined \def \showURL       {\relax}        \fi
\providecommand\bibfield[2]{#2}
\providecommand\bibinfo[2]{#2}
\providecommand\natexlab[1]{#1}
\providecommand\showeprint[2][]{arXiv:#2}

\bibitem[\protect\citeauthoryear{Agarwal, Beygelzimer, Dud{\'\i}k, Langford,
  and Wallach}{Agarwal et~al\mbox{.}}{2018}]%
        {agarwal2018reductions}
\bibfield{author}{\bibinfo{person}{Alekh Agarwal}, \bibinfo{person}{Alina
  Beygelzimer}, \bibinfo{person}{Miroslav Dud{\'\i}k}, \bibinfo{person}{John
  Langford}, {and} \bibinfo{person}{Hanna Wallach}.}
  \bibinfo{year}{2018}\natexlab{}.
\newblock \showarticletitle{A reductions approach to fair classification}. In
  \bibinfo{booktitle}{\emph{International Conference on Machine Learning}}.
  PMLR, \bibinfo{pages}{60--69}.
\newblock


\bibitem[\protect\citeauthoryear{Aggarwal, Lohia, Nagar, Dey, and
  Saha}{Aggarwal et~al\mbox{.}}{2019}]%
        {10.1145/3338906.3338937}
\bibfield{author}{\bibinfo{person}{Aniya Aggarwal}, \bibinfo{person}{Pranay
  Lohia}, \bibinfo{person}{Seema Nagar}, \bibinfo{person}{Kuntal Dey}, {and}
  \bibinfo{person}{Diptikalyan Saha}.} \bibinfo{year}{2019}\natexlab{}.
\newblock \showarticletitle{Black Box Fairness Testing of Machine Learning
  Models}. In \bibinfo{booktitle}{\emph{Proceedings of the 2019 27th ACM Joint
  Meeting on European Software Engineering Conference and Symposium on the
  Foundations of Software Engineering}} \emph{(\bibinfo{series}{ESEC/FSE
  2019})}. \bibinfo{pages}{625–635}.
\newblock
\urldef\tempurl%
\url{https://doi.org/10.1145/3338906.3338937}
\showDOI{\tempurl}


\bibitem[\protect\citeauthoryear{Arcuri and Briand}{Arcuri and Briand}{2014}]%
        {https://doi.org/10.1002/stvr.1486}
\bibfield{author}{\bibinfo{person}{Andrea Arcuri} {and} \bibinfo{person}{Lionel
  Briand}.} \bibinfo{year}{2014}\natexlab{}.
\newblock \showarticletitle{A Hitchhiker's guide to statistical tests for
  assessing randomized algorithms in software engineering}.
\newblock \bibinfo{journal}{\emph{Software Testing, Verification and
  Reliability}} (\bibinfo{year}{2014}), \bibinfo{pages}{219--250}.
\newblock
\urldef\tempurl%
\url{https://doi.org/10.1002/stvr.1486}
\showDOI{\tempurl}


\bibitem[\protect\citeauthoryear{Bellamy, Dey, Hind, Hoffman, Houde, Kannan,
  Lohia, Martino, Mehta, Mojsilovi{\'c}, et~al\mbox{.}}{Bellamy
  et~al\mbox{.}}{2019}]%
        {bellamy2019ai}
\bibfield{author}{\bibinfo{person}{Rachel~KE Bellamy}, \bibinfo{person}{Kuntal
  Dey}, \bibinfo{person}{Michael Hind}, \bibinfo{person}{Samuel~C Hoffman},
  \bibinfo{person}{Stephanie Houde}, \bibinfo{person}{Kalapriya Kannan},
  \bibinfo{person}{Pranay Lohia}, \bibinfo{person}{Jacquelyn Martino},
  \bibinfo{person}{Sameep Mehta}, \bibinfo{person}{Aleksandra Mojsilovi{\'c}},
  {et~al\mbox{.}}} \bibinfo{year}{2019}\natexlab{}.
\newblock \showarticletitle{AI Fairness 360: An extensible toolkit for
  detecting and mitigating algorithmic bias}.
\newblock \bibinfo{journal}{\emph{IBM Journal of Research and Development}}
  \bibinfo{volume}{63}, \bibinfo{number}{4/5} (\bibinfo{year}{2019}),
  \bibinfo{pages}{4--1}.
\newblock


\bibitem[\protect\citeauthoryear{Blumrosen}{Blumrosen}{1978}]%
        {blumrosen1978wage}
\bibfield{author}{\bibinfo{person}{Ruth~G Blumrosen}.}
  \bibinfo{year}{1978}\natexlab{}.
\newblock \showarticletitle{Wage discrimination, job segregation, and the title
  vii of the civil rights act of 1964}.
\newblock \bibinfo{journal}{\emph{U. Mich. JL Reform}}  \bibinfo{volume}{12}
  (\bibinfo{year}{1978}), \bibinfo{pages}{397}.
\newblock


\bibitem[\protect\citeauthoryear{Breiman, Friedman, Olshen, and Stone}{Breiman
  et~al\mbox{.}}{1984}]%
        {Breiman/1984/CART}
\bibfield{author}{\bibinfo{person}{L. Breiman}, \bibinfo{person}{J.H.
  Friedman}, \bibinfo{person}{R.A. Olshen}, {and} \bibinfo{person}{C.I.
  Stone}.} \bibinfo{year}{1984}\natexlab{}.
\newblock \bibinfo{booktitle}{\emph{Classification and regression trees}}.
\newblock \bibinfo{publisher}{Wadsworth: Belmont, CA}.
\newblock


\bibitem[\protect\citeauthoryear{Brun and Meliou}{Brun and Meliou}{2018}]%
        {10.1145/3236024.3264838}
\bibfield{author}{\bibinfo{person}{Yuriy Brun} {and} \bibinfo{person}{Alexandra
  Meliou}.} \bibinfo{year}{2018}\natexlab{}.
\newblock \showarticletitle{Software Fairness} \emph{(\bibinfo{series}{ESEC/FSE
  2018})}. \bibinfo{pages}{754–759}.
\newblock
\urldef\tempurl%
\url{https://doi.org/10.1145/3236024.3264838}
\showURL{%
\tempurl}


\bibitem[\protect\citeauthoryear{Chakraborty, Majumder, Yu, and
  Menzies}{Chakraborty et~al\mbox{.}}{2020}]%
        {chakraborty2020fairway}
\bibfield{author}{\bibinfo{person}{Joymallya Chakraborty},
  \bibinfo{person}{Suvodeep Majumder}, \bibinfo{person}{Zhe Yu}, {and}
  \bibinfo{person}{Tim Menzies}.} \bibinfo{year}{2020}\natexlab{}.
\newblock \showarticletitle{Fairway: a way to build fair ML software}. In
  \bibinfo{booktitle}{\emph{Proceedings of the 28th ACM Joint Meeting on
  European Software Engineering Conference and Symposium on the Foundations of
  Software Engineering}}. \bibinfo{pages}{654--665}.
\newblock


\bibitem[\protect\citeauthoryear{Chakraborty, Majumder, Yu, and
  Menzies}{Chakraborty et~al\mbox{.}}{2021}]%
        {Fairway-tool}
\bibfield{author}{\bibinfo{person}{Joymallya Chakraborty},
  \bibinfo{person}{Suvodeep Majumder}, \bibinfo{person}{Zhe Yu}, {and}
  \bibinfo{person}{Tim Menzies}.} \bibinfo{year}{2021}\natexlab{}.
\newblock \bibinfo{title}{implementation of Fairway}.
\newblock \bibinfo{howpublished}{\url{https://github.com/joymallyac/Fairway}}.
\newblock
\newblock
\shownote{Online.}


\bibitem[\protect\citeauthoryear{Chakraborty, Xia, Fahid, and
  Menzies}{Chakraborty et~al\mbox{.}}{2019}]%
        {chakraborty2019software}
\bibfield{author}{\bibinfo{person}{Joymallya Chakraborty},
  \bibinfo{person}{Tianpei Xia}, \bibinfo{person}{Fahmid~M. Fahid}, {and}
  \bibinfo{person}{Tim Menzies}.} \bibinfo{year}{2019}\natexlab{}.
\newblock \bibinfo{title}{Software Engineering for Fairness: A Case Study with
  Hyperparameter Optimization}.
\newblock
\newblock
\showeprint[late breaking results track (ase 2019)]{1905.05786}


\bibitem[\protect\citeauthoryear{Deb, Pratap, Agarwal, and Meyarivan}{Deb
  et~al\mbox{.}}{2002}]%
        {deb2002fast}
\bibfield{author}{\bibinfo{person}{Kalyanmoy Deb}, \bibinfo{person}{Amrit
  Pratap}, \bibinfo{person}{Sameer Agarwal}, {and} \bibinfo{person}{TAMT
  Meyarivan}.} \bibinfo{year}{2002}\natexlab{}.
\newblock \showarticletitle{A fast and elitist multiobjective genetic
  algorithm: NSGA-II}.
\newblock \bibinfo{journal}{\emph{IEEE transactions on evolutionary
  computation}} \bibinfo{volume}{6}, \bibinfo{number}{2}
  (\bibinfo{year}{2002}), \bibinfo{pages}{182--197}.
\newblock


\bibitem[\protect\citeauthoryear{Deloitte}{Deloitte}{2021}]%
        {deloitte-software}
\bibfield{author}{\bibinfo{person}{Deloitte}.} \bibinfo{year}{2021}\natexlab{}.
\newblock \bibinfo{title}{Better Data, Faster Delivery, Actionable Insights}.
\newblock
  \bibinfo{howpublished}{\url{https://www2.deloitte.com/us/en/pages/deloitte-analytics/solutions/insuresense-insurance-data-analytics-platform-data-management-services.html}}.
\newblock
\newblock
\shownote{Online.}


\bibitem[\protect\citeauthoryear{Dua and Graff}{Dua and Graff}{2017a}]%
        {Dua:2019-census}
\bibfield{author}{\bibinfo{person}{Dheeru Dua} {and} \bibinfo{person}{Casey
  Graff}.} \bibinfo{year}{2017}\natexlab{a}.
\newblock \bibinfo{title}{{UCI} Machine Learning Repository}.
\newblock
\newblock
\urldef\tempurl%
\url{https://archive.ics.uci.edu/ml/datasets/census+income}
\showURL{%
\tempurl}


\bibitem[\protect\citeauthoryear{Dua and Graff}{Dua and Graff}{2017b}]%
        {Dua:2019-credit}
\bibfield{author}{\bibinfo{person}{Dheeru Dua} {and} \bibinfo{person}{Casey
  Graff}.} \bibinfo{year}{2017}\natexlab{b}.
\newblock \bibinfo{title}{{UCI} Machine Learning Repository}.
\newblock
\newblock
\urldef\tempurl%
\url{https://archive.ics.uci.edu/ml/datasets/statlog+(german+credit+data)}
\showURL{%
\tempurl}


\bibitem[\protect\citeauthoryear{Dua and Graff}{Dua and Graff}{2017c}]%
        {Dua:2019-bank}
\bibfield{author}{\bibinfo{person}{Dheeru Dua} {and} \bibinfo{person}{Casey
  Graff}.} \bibinfo{year}{2017}\natexlab{c}.
\newblock \bibinfo{title}{{UCI} Machine Learning Repository}.
\newblock
\newblock
\urldef\tempurl%
\url{https://archive.ics.uci.edu/ml/datasets/bank+marketing}
\showURL{%
\tempurl}


\bibitem[\protect\citeauthoryear{Dwork, Hardt, Pitassi, Reingold, and
  Zemel}{Dwork et~al\mbox{.}}{2012}]%
        {dwork2012fairness}
\bibfield{author}{\bibinfo{person}{Cynthia Dwork}, \bibinfo{person}{Moritz
  Hardt}, \bibinfo{person}{Toniann Pitassi}, \bibinfo{person}{Omer Reingold},
  {and} \bibinfo{person}{Richard Zemel}.} \bibinfo{year}{2012}\natexlab{}.
\newblock \showarticletitle{Fairness through awareness}. In
  \bibinfo{booktitle}{\emph{Proceedings of the 3rd innovations in theoretical
  computer science conference}}. \bibinfo{pages}{214--226}.
\newblock


\bibitem[\protect\citeauthoryear{Galhotra, Brun, and Meliou}{Galhotra
  et~al\mbox{.}}{2017}]%
        {FairnessTesting}
\bibfield{author}{\bibinfo{person}{Sainyam Galhotra}, \bibinfo{person}{Yuriy
  Brun}, {and} \bibinfo{person}{Alexandra Meliou}.}
  \bibinfo{year}{2017}\natexlab{}.
\newblock \showarticletitle{Fairness Testing: Testing Software for
  Discrimination} \emph{(\bibinfo{series}{ESEC/FSE 2017})}.
  \bibinfo{publisher}{Association for Computing Machinery},
  \bibinfo{address}{New York, NY, USA}.
\newblock
\showISBNx{9781450351058}
\urldef\tempurl%
\url{https://doi.org/10.1145/3106237.3106277}
\showDOI{\tempurl}


\bibitem[\protect\citeauthoryear{Hardt, Price, and Srebro}{Hardt
  et~al\mbox{.}}{2016}]%
        {hardt2016equality}
\bibfield{author}{\bibinfo{person}{Moritz Hardt}, \bibinfo{person}{Eric Price},
  {and} \bibinfo{person}{Nati Srebro}.} \bibinfo{year}{2016}\natexlab{}.
\newblock \showarticletitle{Equality of Opportunity in Supervised Learning}. In
  \bibinfo{booktitle}{\emph{NIPS}}.
\newblock


\bibitem[\protect\citeauthoryear{Julia~Angwin and Kirchne}{Julia~Angwin and
  Kirchne}{2021}]%
        {compas-article}
\bibfield{author}{\bibinfo{person}{Surya~Mattu Julia~Angwin, Jeff~Larson} {and}
  \bibinfo{person}{Lauren Kirchne}.} \bibinfo{year}{2021}\natexlab{}.
\newblock \bibinfo{title}{Machine Bias}.
\newblock
  \bibinfo{howpublished}{\url{https://www.propublica.org/article/machine-bias-risk-assessments-in-criminal-sentencing}}.
\newblock
\newblock
\shownote{Online.}


\bibitem[\protect\citeauthoryear{Kamiran, Karim, and Zhang}{Kamiran
  et~al\mbox{.}}{2012}]%
        {6413831}
\bibfield{author}{\bibinfo{person}{Faisal Kamiran}, \bibinfo{person}{Asim
  Karim}, {and} \bibinfo{person}{Xiangliang Zhang}.}
  \bibinfo{year}{2012}\natexlab{}.
\newblock \showarticletitle{Decision Theory for Discrimination-Aware
  Classification}. In \bibinfo{booktitle}{\emph{2012 IEEE 12th International
  Conference on Data Mining}}. \bibinfo{pages}{924--929}.
\newblock
\urldef\tempurl%
\url{https://doi.org/10.1109/ICDM.2012.45}
\showDOI{\tempurl}


\bibitem[\protect\citeauthoryear{Kampmann, Havrikov, Ezekiel, and
  Zeller}{Kampmann et~al\mbox{.}}{2020}]%
        {alhazen}
\bibfield{author}{\bibinfo{person}{Alexander Kampmann},
  \bibinfo{person}{Nikolas Havrikov}, \bibinfo{person}{Soremekun Ezekiel},
  {and} \bibinfo{person}{Andreas Zeller}.} \bibinfo{year}{2020}\natexlab{}.
\newblock \showarticletitle{When does my Program do this? Learning
  Circumstances of Software Behavior} \emph{(\bibinfo{series}{FSE 2020})}.
\newblock


\bibitem[\protect\citeauthoryear{Kuhn, Johnson, et~al\mbox{.}}{Kuhn
  et~al\mbox{.}}{2013}]%
        {kuhn2013applied}
\bibfield{author}{\bibinfo{person}{Max Kuhn}, \bibinfo{person}{Kjell Johnson},
  {et~al\mbox{.}}} \bibinfo{year}{2013}\natexlab{}.
\newblock \bibinfo{booktitle}{\emph{Applied predictive modeling}}.
  Vol.~\bibinfo{volume}{26}.
\newblock \bibinfo{publisher}{Springer}.
\newblock


\bibitem[\protect\citeauthoryear{Nair, Yu, Menzies, Siegmund, and Apel}{Nair
  et~al\mbox{.}}{2020}]%
        {8469102}
\bibfield{author}{\bibinfo{person}{Vivek Nair}, \bibinfo{person}{Zhe Yu},
  \bibinfo{person}{Tim Menzies}, \bibinfo{person}{Norbert Siegmund}, {and}
  \bibinfo{person}{Sven Apel}.} \bibinfo{year}{2020}\natexlab{}.
\newblock \showarticletitle{Finding Faster Configurations Using FLASH}.
\newblock \bibinfo{journal}{\emph{IEEE Transactions on Software Engineering}}
  \bibinfo{volume}{46}, \bibinfo{number}{7} (\bibinfo{year}{2020}),
  \bibinfo{pages}{794--811}.
\newblock
\urldef\tempurl%
\url{https://doi.org/10.1109/TSE.2018.2870895}
\showDOI{\tempurl}


\bibitem[\protect\citeauthoryear{Northpointe}{Northpointe}{2012}]%
        {compas-software}
\bibfield{author}{\bibinfo{person}{Northpointe}.}
  \bibinfo{year}{2012}\natexlab{}.
\newblock \bibinfo{title}{Practitioners Guide to COMPAS}.
\newblock
  \bibinfo{howpublished}{\url{http://www.northpointeinc.com/files/technical_documents/FieldGuide2_081412.pdf}}.
\newblock
\newblock
\shownote{Online.}


\bibitem[\protect\citeauthoryear{O’Whielacronx}{O’Whielacronx}{2018}]%
        {Trace}
\bibfield{author}{\bibinfo{person}{Zooko O’Whielacronx}.}
  \bibinfo{year}{2018}\natexlab{}.
\newblock \bibinfo{title}{A program/module to trace Python program or function
  execution}.
\newblock
  \bibinfo{howpublished}{\url{https://docs.python.org/3/library/trace.html}}.
\newblock
\newblock
\shownote{Online.}


\bibitem[\protect\citeauthoryear{ProPublica}{ProPublica}{2021}]%
        {compas-dataset}
\bibfield{author}{\bibinfo{person}{ProPublica}.}
  \bibinfo{year}{2021}\natexlab{}.
\newblock \bibinfo{title}{Compas Software Ananlysis}.
\newblock
  \bibinfo{howpublished}{\url{https://github.com/propublica/compas-analysis}}.
\newblock
\newblock
\shownote{Online.}


\bibitem[\protect\citeauthoryear{scikit learn}{scikit learn}{2021a}]%
        {Decision-Tree}
\bibfield{author}{\bibinfo{person}{scikit learn}.}
  \bibinfo{year}{2021}\natexlab{a}.
\newblock \bibinfo{title}{Decision Tree Classifier}.
\newblock
  \bibinfo{howpublished}{\url{https://scikit-learn.org/stable/modules/generated/sklearn.tree.DecisionTreeClassifier.html}}.
\newblock
\newblock
\shownote{Online.}


\bibitem[\protect\citeauthoryear{scikit learn}{scikit learn}{2021b}]%
        {Discriminant-Analysis}
\bibfield{author}{\bibinfo{person}{scikit learn}.}
  \bibinfo{year}{2021}\natexlab{b}.
\newblock \bibinfo{title}{Discriminant Analysis}.
\newblock
  \bibinfo{howpublished}{\url{https://scikit-learn.org/stable/modules/lda_qda.html}}.
\newblock
\newblock
\shownote{Online.}


\bibitem[\protect\citeauthoryear{scikit learn}{scikit learn}{2021c}]%
        {logistic-regression}
\bibfield{author}{\bibinfo{person}{scikit learn}.}
  \bibinfo{year}{2021}\natexlab{c}.
\newblock \bibinfo{title}{Logistic Regression}.
\newblock
  \bibinfo{howpublished}{\url{https://scikit-learn.org/stable/modules/generated/sklearn.linear_model.LogisticRegression.html}}.
\newblock
\newblock
\shownote{Online.}


\bibitem[\protect\citeauthoryear{scikit learn}{scikit learn}{2021d}]%
        {random-forest}
\bibfield{author}{\bibinfo{person}{scikit learn}.}
  \bibinfo{year}{2021}\natexlab{d}.
\newblock \bibinfo{title}{Random Forest Regressor}.
\newblock
  \bibinfo{howpublished}{\url{https://scikit-learn.org/stable/modules/generated/sklearn.ensemble.RandomForestRegressor.html}}.
\newblock
\newblock
\shownote{Online.}


\bibitem[\protect\citeauthoryear{scikit learn}{scikit learn}{2021e}]%
        {SVM}
\bibfield{author}{\bibinfo{person}{scikit learn}.}
  \bibinfo{year}{2021}\natexlab{e}.
\newblock \bibinfo{title}{Support Vector Machine}.
\newblock
  \bibinfo{howpublished}{\url{https://scikit-learn.org/stable/modules/generated/sklearn.svm.LinearSVC.html}}.
\newblock
\newblock
\shownote{Online.}


\bibitem[\protect\citeauthoryear{Scism and Maremont}{Scism and
  Maremont}{2010}]%
        {WSJ-insurace}
\bibfield{author}{\bibinfo{person}{Leslie Scism} {and} \bibinfo{person}{Mark
  Maremont}.} \bibinfo{year}{2010}\natexlab{}.
\newblock \bibinfo{title}{Insurers test data profiles to identify risky
  clients.}
\newblock
  \bibinfo{howpublished}{\url{https://www.wsj.com/articles/SB10001424052748704648604575620750998072986}}.
\newblock
\newblock
\shownote{Online.}


\bibitem[\protect\citeauthoryear{Tizpaz{-}Niari, Cern{\'{y}}, Chang, and
  Trivedi}{Tizpaz{-}Niari et~al\mbox{.}}{2018}]%
        {DBLP:conf/aaai/Tizpaz-NiariCCT18}
\bibfield{author}{\bibinfo{person}{Saeid Tizpaz{-}Niari},
  \bibinfo{person}{Pavol Cern{\'{y}}}, \bibinfo{person}{Bor{-}Yuh~Evan Chang},
  {and} \bibinfo{person}{Ashutosh Trivedi}.} \bibinfo{year}{2018}\natexlab{}.
\newblock \showarticletitle{Differential Performance Debugging With
  Discriminant Regression Trees}. In \bibinfo{booktitle}{\emph{Proceedings of
  the Thirty-Second {AAAI} Conference on Artificial Intelligence (AAAI-18)}}.
  \bibinfo{pages}{2468--2475}.
\newblock
\urldef\tempurl%
\url{https://www.aaai.org/ocs/index.php/AAAI/AAAI18/paper/view/16647}
\showURL{%
\tempurl}


\bibitem[\protect\citeauthoryear{Tizpaz-Niari, \v{C}ern\'{y}, and
  Trivedi}{Tizpaz-Niari et~al\mbox{.}}{2020}]%
        {DPFUZZ}
\bibfield{author}{\bibinfo{person}{Saeid Tizpaz-Niari}, \bibinfo{person}{Pavol
  \v{C}ern\'{y}}, {and} \bibinfo{person}{Ashutosh Trivedi}.}
  \bibinfo{year}{2020}\natexlab{}.
\newblock \showarticletitle{Detecting and Understanding Real-World Differential
  Performance Bugs in Machine Learning Libraries}
  \emph{(\bibinfo{series}{ISSTA})}.
\newblock
\urldef\tempurl%
\url{https://doi.org/10.1145/3395363.3404540}
\showDOI{\tempurl}


\bibitem[\protect\citeauthoryear{Udeshi, Arora, and Chattopadhyay}{Udeshi
  et~al\mbox{.}}{2018}]%
        {udeshi2018automated}
\bibfield{author}{\bibinfo{person}{Sakshi Udeshi}, \bibinfo{person}{Pryanshu
  Arora}, {and} \bibinfo{person}{Sudipta Chattopadhyay}.}
  \bibinfo{year}{2018}\natexlab{}.
\newblock \showarticletitle{Automated directed fairness testing}. In
  \bibinfo{booktitle}{\emph{Proceedings of the 33rd ACM/IEEE International
  Conference on Automated Software Engineering}}. \bibinfo{pages}{98--108}.
\newblock


\bibitem[\protect\citeauthoryear{Von~Luxburg}{Von~Luxburg}{2007}]%
        {von2007tutorial}
\bibfield{author}{\bibinfo{person}{Ulrike Von~Luxburg}.}
  \bibinfo{year}{2007}\natexlab{}.
\newblock \showarticletitle{A tutorial on spectral clustering}.
\newblock \bibinfo{journal}{\emph{Statistics and computing}}
  \bibinfo{volume}{17}, \bibinfo{number}{4} (\bibinfo{year}{2007}),
  \bibinfo{pages}{395--416}.
\newblock


\bibitem[\protect\citeauthoryear{Zeller, Gopinath, B{\"o}hme, Fraser, and
  Holler}{Zeller et~al\mbox{.}}{2021}]%
        {fuzzingbook2021}
\bibfield{author}{\bibinfo{person}{Andreas Zeller}, \bibinfo{person}{Rahul
  Gopinath}, \bibinfo{person}{Marcel B{\"o}hme}, \bibinfo{person}{Gordon
  Fraser}, {and} \bibinfo{person}{Christian Holler}.}
  \bibinfo{year}{2021}\natexlab{}.
\newblock \bibinfo{booktitle}{\emph{The Fuzzing Book}}.
\newblock \bibinfo{publisher}{CISPA Helmholtz Center for Information Security}.
\newblock
\urldef\tempurl%
\url{https://www.fuzzingbook.org/}
\showURL{%
\tempurl}
\newblock
\shownote{Retrieved 2021-10-26 15:30:20+02:00.}


\bibitem[\protect\citeauthoryear{Zhang, Lemoine, and Mitchell}{Zhang
  et~al\mbox{.}}{2018}]%
        {zhang2018mitigating}
\bibfield{author}{\bibinfo{person}{Brian~Hu Zhang}, \bibinfo{person}{Blake
  Lemoine}, {and} \bibinfo{person}{Margaret Mitchell}.}
  \bibinfo{year}{2018}\natexlab{}.
\newblock \showarticletitle{Mitigating unwanted biases with adversarial
  learning}. In \bibinfo{booktitle}{\emph{Proceedings of the 2018 AAAI/ACM
  Conference on AI, Ethics, and Society}}. \bibinfo{pages}{335--340}.
\newblock


\bibitem[\protect\citeauthoryear{Zhang and Harman}{Zhang and Harman}{2021}]%
        {DBLP:conf/icse/ZhangH21}
\bibfield{author}{\bibinfo{person}{Jie~M. Zhang} {and} \bibinfo{person}{Mark
  Harman}.} \bibinfo{year}{2021}\natexlab{}.
\newblock \showarticletitle{"Ignorance and Prejudice" in Software Fairness}. In
  \bibinfo{booktitle}{\emph{43rd {IEEE/ACM} International Conference on
  Software Engineering, {ICSE} 2021, Madrid, Spain, 22-30 May 2021}}.
  \bibinfo{publisher}{{IEEE}}, \bibinfo{pages}{1436--1447}.
\newblock
\urldef\tempurl%
\url{https://doi.org/10.1109/ICSE43902.2021.00129}
\showDOI{\tempurl}


\bibitem[\protect\citeauthoryear{Zhang, Harman, Ma, and Liu}{Zhang
  et~al\mbox{.}}{2020a}]%
        {zhang2020machine}
\bibfield{author}{\bibinfo{person}{Jie~M Zhang}, \bibinfo{person}{Mark Harman},
  \bibinfo{person}{Lei Ma}, {and} \bibinfo{person}{Yang Liu}.}
  \bibinfo{year}{2020}\natexlab{a}.
\newblock \showarticletitle{Machine learning testing: Survey, landscapes and
  horizons}.
\newblock \bibinfo{journal}{\emph{IEEE Transactions on Software Engineering}}
  (\bibinfo{year}{2020}).
\newblock


\bibitem[\protect\citeauthoryear{Zhang, Wang, Sun, Dong, Wang, Wang, Dong, and
  Dai}{Zhang et~al\mbox{.}}{2020b}]%
        {ADF}
\bibfield{author}{\bibinfo{person}{Peixin Zhang}, \bibinfo{person}{Jingyi
  Wang}, \bibinfo{person}{Jun Sun}, \bibinfo{person}{Guoliang Dong},
  \bibinfo{person}{Xinyu Wang}, \bibinfo{person}{Xingen Wang},
  \bibinfo{person}{Jin~Song Dong}, {and} \bibinfo{person}{Ting Dai}.}
  \bibinfo{year}{2020}\natexlab{b}.
\newblock \showarticletitle{White-Box Fairness Testing through Adversarial
  Sampling} \emph{(\bibinfo{series}{ICSE '20})}.
  \bibinfo{publisher}{Association for Computing Machinery},
  \bibinfo{address}{New York, NY, USA}.
\newblock
\showISBNx{9781450371216}
\urldef\tempurl%
\url{https://doi.org/10.1145/3377811.3380331}
\showDOI{\tempurl}


\end{thebibliography}
